\documentclass[manuscript,screen, nonacm]{acmart}

\usepackage{graphicx}

\usepackage{colortbl}
\usepackage{lscape}

\AtBeginDocument{%
  }

\begin{document} 
\newcommand\Afsaneh[1]{\textbf{[\textcolor{purple}{#1}\textemdash\scshape{Afsaneh}]}}
\newcommand\Jordyn[1]{\textbf{[\textcolor{blue}{#1}\textemdash\scshape{Jordyn}]}} 

\title[Epilepsy Online Social Support]{Epilepsy Online Social Support: Characterizing Topics and Challenges Shared in the r/Epilepsy Community}


\author{Jessica Y. Medina}
\email{jym29@drexel.edu}
\orcid{0000-0003-3420-7424}
\affiliation{%
  \institution{Drexel University}
  \city{Philadelphia}
  \state{Pennsylvania}
  \country{USA}
} 

\author{Jordyn Young}
\email{jordyny@umich.edu}
\orcid{0000-0002-1565-0642}
\affiliation{%
  \institution{University of Michigan}
  \city{Ann Arbor}
  \state{Michigan}
  \country{USA}
} 

\author{Aehong Min}
\email{aehongm@uci.edu}
\orcid{0000-0002-3790-2126}
\affiliation{%
  \institution{University of California, Irvine}
  \city{Irvine}
  \state{California}
  \country{USA}
}

\author{Patrick C. Shih}
\email{patshih@iu.edu}
\orcid{0000-0003-2460-0468}
\affiliation{%
  \institution{Indiana University}
  \city{Bloomington}
  \state{Indiana}
  \country{USA}
} 

\author{Wendy R. Miller}
\email{wrtruebl@iu.edu}
\orcid{0000-0003-1779-7290}
\affiliation{%
  \institution{Indiana University}
  \city{Bloomington}
  \state{Indiana}
  \country{USA}
} 

\author{Afsaneh Razi}
\email{ar3882@drexel.edu}
\orcid{0000-0001-5829-8004}
\affiliation{%
  \institution{Drexel University}
  \city{Philadelphia}
  \state{Pennsylvania}
  \country{USA}
}

\begin{abstract}

Epilepsy is one of the most common neurological conditions, and people living with epilepsy (PLWE) often use social media as a resource. However, a comprehensive understanding of the topics represented in epilepsy-specific communities where PLWE may be more honest is essential to designing better technologies to address epilepsy self-management. 
To understand the main topics and concerns of PLWE, we collected 23,944 r/Epilepsy subreddit posts and performed topic modeling, thematic, and psycho-linguistic analyses. We found five major themes for those topics: symptoms and triggers (e.g., mental health and memory, sleep/nocturnal, and photosensitivity), treatment and healthcare experience (e.g., medication, understanding epilepsy), daily functions ( e.g., perceived level of independence and finances), seizure activity (e.g., auras and ictal symptoms), and support for PLWE (assisting PLWE and support for PLWE). We highlight the top psycho-linguistic characteristics of posts across different topics. 
Our contributions include providing an understanding of the challenges of an online epilepsy community and their social support needs, and implications for designing technologies.  



\end{abstract}






\maketitle 
\section{Introduction}

Epilepsy is a common neurological condition affecting more than 1\% of the global population and approximately 3.4 million people in the United States \cite{WorldHealthOrganization2019, CDC2020a}. Beyond seizure and medication management, people living with epilepsy (PLWE) often experience stigma, emotional and mental health challenges, and reduced quality of life \cite{England2012, Miller2014, min2021just}. 

Social media platforms have become important spaces where people with chronic illnesses seek informational and emotional support \cite{park2020health, Sannon-invisiblechronicillness}. These platforms provide large-scale, diverse datasets that can reveal health-related concerns and experiences not easily captured through traditional methods such as surveys or interviews \cite{priest2016finding}. Prior work has explored online support-seeking behaviors for mental health, chronic and stigmatized conditions \cite{Balani2015, Guimar2021, Chancellor2019, Progga2023,Razi2020lets}. 

Existing epilepsy-related social media studies have largely relied on keyword searches or smaller condition-specific communities \cite{fazekas_insights_2021, meng_social_2017, he_understanding_2019}. While valuable, these approaches may overlook broader lived experiences and discussions that do not explicitly reference epilepsy. In contrast, Reddit’s \textit{r/Epilepsy} community contains over 59{,}000 members, offering a large and diverse space for PLWE to discuss their experiences online.

To better understand these discussions, we addressed the following research questions:

\textbf{RQ1:} \textit{What are the main topics and challenges discussed in a popular online epilepsy-focused community?}

\textbf{RQ2:} \textit{What are the top psycho-linguistic characteristics of the topics discussed in the posts?}

We analyzed 23{,}944 posts from \textit{r/Epilepsy} using Latent Dirichlet Allocation (LDA) topic modeling, thematic analysis, and Linguistic Inquiry and Word Count (LIWC). We identified 15 discussion topics grouped into five overarching themes: Symptoms and Triggers, Treatment and Healthcare Experience, Daily Functions, Seizure Activity, and Support for PLWE. 

This work contributes by: (1) providing a large-scale understanding of the concerns discussed by PLWE online, (2) identifying psycho-linguistic characteristics associated with different discussion topics, and (3) offering implications for designing technology-supported self-management and peer-support interventions for PLWE.

\section{Background} 

\subsection{Epilepsy Health Condition and Self-Management}

Epilepsy is one of the most common neurological disorders that causes recurrent seizures~\cite{WorldHealthOrganization2019, CDC2020a}. 
A seizure is an abrupt, unforeseeable, and uncontrollable surge of electrical activity within the brain, which can endure for an extended period. Seizure symptoms can vary significantly; some PLWE display observable signs, like falling or trembling, whereas others' symptoms are concealed and imperceptible, such as impaired vision and momentary loss of awareness \cite{Mayo2021}. Furthermore, PLWE may struggle to recall when the seizure occurred \cite{Mayo2021}. Epilepsy can also lead to other serious health problems and increase the risk of premature death \cite{WorldHealthOrganization2019, CDC2020a, Watila2018}. 

Epilepsy has diverse effects on the health and daily lives of PLWE and the people around them. Although effective medication and treatment options enable most PLWE to manage their seizures \cite{Kwan2000}, epilepsy poses significant physical, psychological, social, and financial burdens on both PLWE and their caregivers \cite{DeBoer2008}. PLWE often experience physical challenges such as limited mobility and fatigue, which can result in reduced focus, performance, and various limitations in their professional and personal lives \cite{Unger2009}. There are additional risks associated with medication side effects, cognitive impairments, memory problems, social isolation, and the stigma surrounding epilepsy \cite{England2012, mahrer2013quality, ShettyA2017, DeBoer2008, Jacoby2002}. These factors, along with psychological and emotional issues like depression, anxiety, and frustration arising from stigma, public misconceptions, and uncertainty about experiencing seizures, have a negative impact on the lives of PLWE and their caregivers \cite{Freestone2017}. PLWE frequently face feelings of loneliness and worry about how others will perceive them and react to their seizures in public. They may also develop a more negative perception of the changes epilepsy has brought to their lives, such as disruptions in their ability to sleep well, which can potentially trigger further seizures \cite{DeBoer2008, miller2016perceived, Beyenburg2005}. PLWE also encounters challenges related to receiving low-quality care, complex care coordination, the risk of sudden unexpected death in epilepsy (SUDEP), and unstable employment \cite{England2012, Miller2014}.

People with chronic conditions such as epilepsy are responsible for their daily management, monitoring for behavior changes and emotional adjustments, and reporting trends and tempos of their condition \cite{holman2004patient}. As a result, developing self-management skills is an essential tool for optimum health in PLWE \cite{buelow2000self}. Previous researchers have focused on chronic condition self-management programs and efficacy \cite{lorig2001chronic, lorig2001effect, lorig1999evidence} as well as frameworks and concepts \cite{mills2017toward, miller_chronic_2015, novak2013approaches, van2019delineating}. Miller et al. highlighted the complex process of chronic condition self-management. They proposed a hybrid concept analysis of chronic condition self-management based on the literature at the time and patient-based descriptions \cite{miller_chronic_2015}. This framework highlights that chronic condition self-management incorporates multidimensional strategies because it requires the individual to incorporate intrapersonal, interpersonal, and environmental systems \cite{miller_chronic_2015}.

\subsection{Epilepsy in HCI and Computing} 
Previous research in HCI studied how PLWE and caregivers manage epilepsy-related information and the challenges they encountered \cite{min2023understanding}. In addition to epilepsy-related information, researchers have analyzed the prevalence of seizure-inducing GIFs across social media platforms \cite{South2021GIFs}. South et al. also focused on photosensitivity warnings \cite{South2023Warnings} and photosensitivity in virtual reality \cite{South2024}. Research in HCI on technology in epilepsy has predominantly revolved around creating and evaluating technologies capable of detecting and alerting individuals about seizures \cite{Petroulaki2020}. For example, studies have evaluated methods such as deep learning to discuss how these technologies can improve the prediction and detection of epilepsy seizures \cite{Hussein2020, Hossain2019}. In addition to detection and prediction, researchers such as Bruno et. al highlighted the vast interest in using wearable technology \cite{bruno2018wearable}. Furthermore, other HCI studies have investigated the implementation of intelligent decision systems to improve diagnosis \cite{Tahir2019, Marathe2021}, systems for prescribing medication dosages based on individual pharmacological responses \cite{Helling2019}, neurofeedback training games aimed at reducing seizures \cite{Reinschluessel2016game}, and monitoring systems for predicting SUDEP \cite{Yue2020}. Despite the extensive exploration of these assistive systems for PLWE, the epilepsy literature needs more research to investigate the lived experience of PLWE through their online help-seeking and information-sharing practices to address the challenges they encounter in daily life.  Understanding PLWE's online support-seeking and information-sharing behavior can further build on existing research and aid future researchers in determining other potential areas in the lives of PLWE that can benefit from the design and implementation of systems and interventions such as these. 

Research in the area of computing focused on technology for seizure detection and prediction \cite{hussein2020AugmentingDLAdversarial, aziz2025UnobtrusiveLightweightEarworn}. 
Previous research on technologies for epilepsy expands into various applications such as medicine \cite{marathe2020seizure, shegog2020digital}, machine learning applications \cite{rasheed2020machine, Jia2024}, wearable technology \cite{Rashid2024, brinkmann2021seizure}, and AI systems \cite{Yu2024, Cheng2024}. Medical technologies for epilepsy focused on how technological progress is impacting the landscape of epilepsy surgery \cite{dorfer2020technology}, the use of medical technology in epilepsy diagnosis \cite{ong2022medical}, as well as devices for seizure control \cite{stacey2008technology}. Researchers in computing focused on leveraging machine learning for seizure prediction, AI for detection and classification, and wearable technologies. Usman et al. proposed a model that predicts seizures with sufficient time before the onset of seizures, with an average prediction time of 23.6 minutes \cite{usman2017epileptic}. 
Concerning wearable technology, researchers focused on its ability to predict seizures.
Stirling et al. incorporated machine learning models to analyze heart rate as a biomarker alongside other wearable signals to forecast high and low seizure risk periods \cite{stirling2021forecasting}. 
Alongside this, previous researchers reviewed e-tools \cite{shegog2013managing}, digital tools \cite{hixson_digital_2020, shegog2020digital} and mobile applications for epilepsy care and self-management such as Managing Epilepsy Well (MEW) \cite{escoffery_review_2018, alzamanan_self-management_2021, choi2021impact, ernawati2024effect}.
Additionally, researchers discovered that mobile applications for epilepsy incorporated features such as seizure logging \cite{abreu2022mobile}, medication \cite{choi2021impact}, and treatment management \cite{escoffery_review_2018}.

\subsection{Epilepsy in Online Communities and Social Media}
Online communities and social media have been shown to have a positive impact on one's well-being and health \cite{Hong2013, Moorhead2013, Hayes2016}. Several platforms have been studied to understand perceptions of PLWE and explore epilepsy-relevant content on social media and online communities and its impacts \cite{de_la_loge_patientslikeme_2016, wicks_perceived_2012, Wong2013, baxendale_epilepsy_2021}. For instance, He et al.~\cite{he_understanding_2019} focused on identifying epilepsy treatment-related themes from patients' posts on three online patient support groups: Mayo Clinic, Coping-with-epilepsy.com, and Epilepsy.com. They discovered that patients sought help and information about medical treatments from other posters and shared their own treatment experiences. 
De la Loge et al.~\cite{de_la_loge_patientslikeme_2016} conducted a retrospective analysis on a dataset from PatientsLikeMe to characterize the profile of users and their conditions and identify factors predictive of poor health-related quality of life. Building on these studies, our work focuses on further learning of treatment-related themes and broader epilepsy-related themes within a specific community. This allows further insight into other concerns that can impact quality of life, allowing future work to take a more holistic approach when designing for PLWE. When it comes to social media, some studies have focused on the types and content of epilepsy-related Facebook and Twitter accounts and revealed that they consist of foundations or support groups sharing information about medications and correcting misconceptions \cite{meng_social_2017, MCNEIL2012127}. In contrast, others focused on analyzing the digital behavior of patients deceased due to SUDEP \cite{wood2022small}. Researchers also explored epilepsy on social media presented in other languages \cite{karadvzic2022epilepsy, alsalem2021epilepsy}. While research on people with chronic illness has focused on diverse social, practical, and political support-seeking and information-sharing behaviors in online peer-to-peer interactions \cite{kingod_online_2017, Patel2015, Sendra2020, Sannon2019}, few research have systematically explored PLWE support-seeking behaviors on social media and online communities. 

Reddit is a social media platform comprising numerous forums called 'subreddits' \cite{Reddit2023}. Data from these subreddits have been studied using various methods (e.g., text mining, NLP, content analysis) to understand users' common topics and communication behaviors, such as support-seeking, crowd-sourcing, and disclosing in diverse contexts \cite{Sannon2019, Ernala2022_hospitalization, Tejaswini2022}. Some studies have analyzed health-relevant topics on Reddit, such as chronic conditions, depression, Gout, inflammatory bowel disease \cite{Sannon2019, Ernala2022_hospitalization, Tejaswini2022, Liu2023_detecting, Kim2023_supporters, Rohde2021_IBD}.
There are also several subreddits related to neurological conditions on Reddit. Researchers investigated 605 posts in 15 subreddits related to neurological conditions, and they identified that users commonly asked for advice about treatments \cite{To2023}. The number of members in the subreddit \textit{r/Epilepsy} is the highest among these subreddits \cite{To2023, Reddit2023}. Due to the volume of users and activity, research on \textit{r/Epilepsy} would provide more valid data to understand PLWE's support-seeking and information-sharing behaviors as well as to identify the PLWE's challenges that should be addressed. 

Overall, while previous work analyzed epilepsy-focused online communities, these communities are less popular than other social media platforms, such as Reddit. Work on social media for epilepsy typically depended on keyword searches such as seizure, which can limit content, since epilepsy is more than seizures. Additionally, no work has analyzed the psycholinguistic features of what is discussed among PLWE. Conducting this study is important because it addresses these gaps by broadening the scope of epilepsy-related research beyond seizure-centric discussions and beyond niche forums, capturing more naturalistic and diverse conversations occurring on widely used platforms. Understanding the psycholinguistic features of these discussions can provide insight into the emotional burden, stigma, coping strategies, and unmet needs experienced by PLWE. This knowledge can inform patient-centered interventions, enhance clinicians' awareness of lived experiences, guide public health communication strategies, and ultimately support more holistic epilepsy care that addresses the psychosocial as well as clinical dimensions of the condition.

\section{Methods}
\subsection{Data Collection and Ethical Considerations}
We collected and analyzed data from an active online community of PLWE: the r/Epilepsy subreddit. Subreddits are communities within the social media platform Reddit ~\cite{Reddit2023}; with over 500 million users on the platform, Reddit is ideal for studying text-based online activity from specific communities because the datasets are large and present various perspectives \cite{Progga2023, gao2021understanding}. The r/Epilepsy subreddit specifically has more than 59,000 members. While there are subreddits for discussing seizures or neurology (e.g., r/seizures \footnote{https://www.reddit.com/r/seizures/}, r/seizure \footnote{https://www.reddit.com/r/Seizure/}, and r/neurology \footnote{https://www.reddit.com/r/neurology/}), which are key terms but we chose not to include them in the present study. This is because seizures can be caused by many conditions other than epilepsy, such as metabolic problems, psychogenic seizures, and electrolyte problems, and our main focus was on PLWE and epilepsy-related concerns \footnote{\url{https://my.clevelandclinic.org/health/diseases/22789-seizure}}. While neurology encompasses other conditions such as stroke, ALS, migraines, Alzheimer's, Parkinson's, and multiple sclerosis \footnote{https://www.aan.com/tools-resources/what-is-a-neurologist}. We used PMAW~\footnote{\url{https://reddit-api.readthedocs.io/en/latest/}}, a Python wrapper for the Pushshift API, to collect all the posts that were published in r/Epilepsy from its initial creation in 2010 up until March 23rd, 2023 (n = 26,844). 

For ethical reasons, although the data collected are considered public and determined as non-human subject research by our Institutional Review Board (IRB), we proceeded with utmost care to ensure users' privacy in multiple ways. First, we anonymized the authors of posts in the dataset by using the hashlib module~\footnote{\url{https://docs.python.org/3/library/hashlib.html}}. Second, we paraphrased all quotes included in this paper to protect users' confidentiality, as previous research recommended disguising authors and protecting them from potentially being reversed searched \cite{bruckman2002studying}. Third, we confirmed that at least one author on this paper has lived experience with epilepsy and at least one does not, which guided the research process to include a combination of in- and out-group perspectives.
Finally, we removed posts deleted or removed by users since the time of posting (n = 2,904) to respect their decisions about their content.
Our final dataset of 23,944 posts was used for further analysis. 

\subsection{Topic Modeling and Subsequent Analysis}  
We conducted a mixed-method approach by implementing topic modeling to identify clusters of topics \cite{churchill_evolution_2022}, thematic analysis to identify, analyze, and report recurring content within each topic, and LIWC to characterize sentiment and linguistics features. 
While our data extraction included flairs for 6,276 of the 26,844 posts, we did not include it in our base categorization due to the majority of posts not having text flairs and the nature of topic modeling not requiring preliminary categorization. 

To pick the right topic modeling for our task, we compared results from recent BERTopic~\cite{2022bertopic} and classic LDA~\cite{jelodar2019latent}. 
%
To identify the optimal number of topics for LDA, we calculated coherence scores for possible topic counts and found the optimal number of topics to be n = 15, with a coherence score of 0.4738. We discovered that after 15 topics, the coherence score continued to decrease steadily by checking up to 32 topic counts.
We also implemented the BERTopic model with the topic number parameter set to n = 15 to better compare both LDA and BERTopic. 
While we acknowledge that the BERTopic is more advanced, it requires additional steps such as dimensionality reduction, which adds more complexity and subjectivity that may not necessarily make the topics more interpretable. As the contribution of our paper was to understand the topics that people with epilepsy seek support online, we prioritized methods that yield topics with greater human interpretability~\cite{egger2022topic}.
Ultimately, we found the results of the LDA model sufficient and more interpretable due to the model's results showing each of the document (post) probability distribution to its assigned topic, which aided the researchers to better identify topic names and themes in the subsequent analysis. 
Within each resulting topic, there was a grouping of similar words to indicate which words closely belonged with each other. 

To gain a deeper understanding of the LDA topics, we conducted a thematic analysis~\cite{thematic2006}. The topics from the LDA guided our deductive thematic analysis in conceptualizing the themes. Our procedure for the thematic analysis included an initial round in which two researchers reviewed the top 10 words for each topic and the top posts sorted by the probability of belonging to the assigned topic to gather preliminary themes. This was followed by rounds of discussion between the first and second authors. We discovered that the topic definition was apparent before reviewing the 100th post, and we reached topical saturation. Therefore, we normalized the number of posts for review for the rest of the topics to 100. This number also encompassed posts with less than a 0.50 probability distribution within smaller topics. Since our thematic analysis was deductive based on the topics provided by LDA, the use of a codebook and the calculation of an IRR were unnecessary~\cite{mcdonald2019reliability}. 
The team frequently met to finalize and describe the themes and topics and resolve any disagreements for the final decision.
After several team discussions, we categorized topics into overarching themes based on similarities found among the topics, as displayed in Table 1. 

\subsubsection{Psycho-linguistic Analysis}
We fed the posts from each topic through LIWC-22 to evaluate each post within our dataset by cross-referencing with the LIWC-22 dictionary \cite{LIWC2022}. The LIWC-22 analysis provided insight into linguistic characteristics of posts, such as the use of pronouns and tense, as well as psychological processes, such as affective, social, and cognitive processes \cite{Matthews2015, Che2018}. 
We present the mean of each LIWC category per topic in Appendix \ref{section-affect}. When we describe the topics in the results, we report whether they had the highest mean of a LIWC category compared with other topics. This will help highlight the psycho-linguistic characteristics each topic has.




\section{Results}  

We present the results of our topic modeling, which revealed 15 unique topics in our data, ordered by prevalence (visualized in Table 1). 
Please note that the topic numbers in the LDA output do not represent any ordinal ranking.

We also report the means of the LIWC categories if that topic had the highest mean of the LIWC category to highlight its psycho-linguistic characteristics.  
To aid interpretation, we underline representative words and phrases in example posts that correspond to LIWC-22 categories with a higher mean score(s).



\begin{table}[t!]
\small
\caption{Themes, topics, top words and topic distribution}
\resizebox{\textwidth}{!}{%
\begin{tabular}{|lll|}
\hline
\multicolumn{1}{|c|}{\textbf{Topic \# - Topic Description}}                                           & \multicolumn{1}{c|}{\textbf{Top words}}                                                                                                                         & \multicolumn{1}{c|}{\textbf{Distribution}} \\ \hline
\multicolumn{3}{|c|}{\textit{\textbf{Theme: Symptoms and Triggers (31.56\%, N = 7,560)}}}                                                                                                                                                                                                                            \\ \hline
\multicolumn{1}{|l|}{1 - Mental health and memory}                                                    & \multicolumn{1}{l|}{\begin{tabular}[c]{@{}l@{}}Feel, know, thing, want, think, people, bad, \\ time, life, tell\end{tabular}}                                   & 18.25\%, N = 4371                          \\ \hline
\multicolumn{1}{|l|}{0 - Sleep/Nocturnal}                                                             & \multicolumn{1}{l|}{\begin{tabular}[c]{@{}l@{}}Seizure, sleep, night, day, time, wake, go, \\ hour, happen, morning,\end{tabular}}                              & 11.96\%, N = 2865                          \\ \hline
\multicolumn{1}{|l|}{2 - Photosensitivity}                                                            & \multicolumn{1}{l|}{\begin{tabular}[c]{@{}l@{}}Epilepsy, game, seizure, play, light, epileptic, \\ movie, trigger, flashing light, photosensitive\end{tabular}} & 1.35\%, N = 324                            \\ \hline
\multicolumn{3}{|c|}{\textit{\textbf{Theme: Treatment and Healthcare Experience (28.01\%, N = 6707)}}}                                                                                                                                                                                                               \\ \hline
\multicolumn{1}{|l|}{4 - Medication}                                                                  & \multicolumn{1}{l|}{\begin{tabular}[c]{@{}l@{}}Take, effect, medication, day, Keppra, Lamictal, \\ med, start, experience, week\end{tabular}}                   & 10.94\%, N = 2620                          \\ \hline
\multicolumn{1}{|l|}{11 - Understanding epilepsy}                                                     & \multicolumn{1}{l|}{\begin{tabular}[c]{@{}l@{}}Seizure, epilepsy, diagnose, year, eeg, experience, \\ diagnosis, symptom, absence, cause\end{tabular}}          & 7.88\%, N = 1886                           \\ \hline
\multicolumn{1}{|l|}{7 - Healthcare experience}                                                       & \multicolumn{1}{l|}{\begin{tabular}[c]{@{}l@{}}Doctor, say, tell, neurologist, ask, test, go, \\ question, appointment, know\end{tabular}}                      & 6.00\%, N = 1437                           \\ \hline
\multicolumn{1}{|l|}{12 - Medical research/treatment}                                                 & \multicolumn{1}{l|}{\begin{tabular}[c]{@{}l@{}}Surgery, brain, seizure, epilepsy, patient, \\ treatment, risk, study, pregnancy, option,\end{tabular}}          & 3.19\%, N = 764                            \\ \hline
\multicolumn{3}{|c|}{\textit{\textbf{Theme: Daily Function (22.04\%, N = 5,277)}}}                                                                                                                                                                                                                                   \\ \hline
\multicolumn{1}{|l|}{\begin{tabular}[c]{@{}l@{}}14 - Perceived level of \\ Independence\end{tabular}} & \multicolumn{1}{l|}{\begin{tabular}[c]{@{}l@{}}Seizure, year, month, drive, time, get, go, \\ medication, epilepsy, day\end{tabular}}                           & 18.46\%, N = 4420                          \\ \hline
\multicolumn{1}{|l|}{8 - Finances}                                                                    & \multicolumn{1}{l|}{\begin{tabular}[c]{@{}l@{}}Work, job, insurance, need, pay, disability, \\ money, health, company, cost\end{tabular}}                       & 2.25\%, N = 539                            \\ \hline
\multicolumn{1}{|l|}{13 - Lifestyle choices}                                                          & \multicolumn{1}{l|}{\begin{tabular}[c]{@{}l@{}}Drink, eat, alcohol, food, diet, weight, \\ smoke, high, day, exercise\end{tabular}}                             & 1.33\%, N = 318                            \\ \hline
\multicolumn{3}{|c|}{\textit{\textbf{Theme: Seizure Activity (12.63\%, N = 2,111)}}}                                                                                                                                                                                                                                 \\ \hline
\multicolumn{1}{|l|}{10 - Auras}                                                                      & \multicolumn{1}{l|}{\begin{tabular}[c]{@{}l@{}}Feel, seizure, happen, start, experience, think, \\ time, feeling, aura, weird\end{tabular}}                     & 9.12\%, N = 1271                           \\ \hline
\multicolumn{1}{|l|}{3 - Ictal}                                                                       & \multicolumn{1}{l|}{\begin{tabular}[c]{@{}l@{}}Pain, head, eye, face, leave, body, \\ arm, hand, fall, right\end{tabular}}                                      & 2.67\%, N = 639                            \\ \hline
\multicolumn{1}{|l|}{\begin{tabular}[c]{@{}l@{}}5 - Seizure monitoring\\ /alert\end{tabular}}         & \multicolumn{1}{l|}{\begin{tabular}[c]{@{}l@{}}Video, watch, com, device, eeg, app, \\ wear, monitor, article, use\end{tabular}}                                & 0.84\%, N = 201                            \\ \hline
\multicolumn{3}{|c|}{\textit{\textbf{Theme: Support for PLWE (5.76\%, N = 1379)}}}                                                                                                                                                                                                                                   \\ \hline
\multicolumn{1}{|l|}{9 - Assisting PLWE}                                                              & \multicolumn{1}{l|}{\begin{tabular}[c]{@{}l@{}}Epilepsy, people, help, share, experience, post, \\ thank, read, story, class\end{tabular}}                      & 5.31\%, N = 1271                           \\ \hline
\multicolumn{1}{|l|}{6 - Support for PLWE}                                                            & \multicolumn{1}{l|}{\begin{tabular}[c]{@{}l@{}}Dog, sudep, team, awareness, death, walk, \\ train, raise, police, alert\end{tabular}}                           & 0.45\%, N = 108                            \\ \hline
\end{tabular}%
}
\end{table}

\subsection{Symptoms and Triggers} 
Symptoms refer to physical and mental indicators of a condition. Triggers are factors that can provoke seizures in those who are susceptible and make it more likely for PLWE to have a seizure~\cite{balamurugan2013perceived}. We found that these topics are intertwined because of their interchangeability between being discussed as a symptom or a trigger. 
Mental health, memory issues, and sleep topics included the context of both as symptoms and triggers. However, Photosensitivity, which, for a small percentage of PLWE, is a severe trigger.
\subsubsection{Mental Health and Memory}  
We observed a critical topic wherein PLWE discussed depression, stress, anxiety, and issues related to memory, showing the impacts of epilepsy on both the physical and mental health of PLWE. This topic is the largest as it includes emotional, psychological, and social well-being for PLWE. Our LIWC-22 analysis revealed that posts about this topic contained stronger, more emotional words, such as negative emotions, anger, and sadness, than other topics. Posts about this topic contain one of the highest means of 5.98 in affect, with 2.86 in negative tones and 2.76 in emotion.

We noticed how Redditors considered this subreddit a safe place to share their experiences with epilepsy and openly discuss their emotions:
\emph{``People tell me to be\underline{positive} all the time and don't let me rant about epilepsy! I want to rant and don't want to be \underline{positive} when it fucks up life. Please rant here about how much it fucking sucks''}
This post invited others in the community to share their feelings, particularly negative feelings. In this case, we observed the Redditor's post closer to a perceived bleak reality of living with epilepsy, which further corroborates the results of our LIWC-22 analysis. See Appendix \ref{section-affect}.

While many posts lean toward a gloomier side, including venting about living with epilepsy, others showcased how grateful PLWE were for the r/Epilepsy subreddit, regardless of how much content they contribute: \emph{``Thank you! I \underline{feel so alone}, but this sub has been helpful. I feel less like I'm losing my mind. Even if I don't comment, thanks.''}
Similar to the previous post, this post highlighted how important the r/Epilepsy subreddit is to members of the epilepsy community. Even though the author of the previous post said they are not a particularly active member of the community, they still express gratitude and appreciation. This gratitude for the community represents an optimistic side of how PLWE handle their feelings with epilepsy.

In terms of psychological health, we've noticed that PLWE also worry about stress, anxiety, and depression. Redditors pointed out that their anxiety stems from the fear of when the next seizure will happen. For example, the following Redditor expressed this concern and explained how it is imperative for them to simultaneously prioritize their mental and physical health because they are heavily intertwined: 

\emph{``How do y'all stay positive? Epilepsy makes me \underline{anxious.} Lately, I've been \underline{feeling anxious and demoralized} even though my seizures are controlled with medication, and I'm \underline{worried} that I'm going to have another seizure. My biggest triggers are lack of sleep and \underline{stress.} Can anyone relate?''}

This post demonstrated that some PLWE experience stress concerning health, even with proper self-management. As the Redditor noted, this contributes to feelings of dread and demoralization, which aligns with medical knowledge that depression, anxiety, and obsessive-compulsive disorder are significant comorbidities for PLWE. 

In addition to mental health, we learned that some PLWE expressed concern over cognitive functions, specifically memory. Some Redditors posted their personal experiences with memory issues, while others crowdsource information and support on this matter. This 26-year-old Redditor discussed their memory and mental issues, stating that their memory issues cause them to feel as if they are much older.

\emph{``I'm 26 but my memory and general mental issues are so \underline{bad} I feel like I should be being pushed around in a wheelchair lolz seriously I laugh but it sucks cause my young adult life is wasting away cause I have  issues with memory and focus.''}

This Redditor expressed disheartenment, additionally, we saw other posts made by PLWE looking for both solutions and validation for their memory issues. As this Redditor noted, they experienced an emotional response to their severe memory loss and were looking for both validation and advice to combat this problem: 
\emph{``Does anyone else have memory loss? My family reminds me of places we've been and things I've done. \underline{It's really getting to me} that I don't remember a damn thing. How did you figure out a way to cope with memory problems?''}
Though memory concerns were not the prominent point of discussion, their implications have proved to cause a negative emotional response as seen above.

\subsubsection{Sleep and Nocturnal}
A prominent topic of discussion on r/Epilepsy was sleep, which is closely linked to epilepsy. According to the LIWC-22 analysis, this topic has one of the highest means of 73.49 in the linguistic category, with the personal pronoun 'I' (mean = 8.92) being used the most. In addition to linguistics, this topic had the highest mean (0.36) in the category of fatigue. Many Redditors discussed sleep as a potential trigger for seizures, a postictal symptom, and as a side effect of medications. Redditors also discussed their nocturnal cases of epilepsy activities, accounts of epilepsy, and how anti-epilepsy drugs affect their sleep. 

Some Redditors described how their sleep is affected after a seizure activity.  
\emph{``\underline{I} slept for 18 hours after my seizure yesterday and \underline{I} am \underline{exhausted}. I'm going on holiday tomorrow and am exhausted.''}
The Redditor discussed that after having a seizure, they overslept and still felt exhausted, which is described as a post-ictal symptom of extreme fatigue. This post-ictal symptom can disrupt sleeping patterns in PLWE. 

PLWE also discussed concerns about sleep deprivation and insomnia. which has been linked with triggering seizure activity, and some Redditors worry about how that may affect them the next day: 
\emph{``Is \underline{lack of sleep} a trigger for anyone else? \underline{I've} been lying in bed trying to fall \underline{asleep} for over 3 hours. It's after 2 am. Tomorrow is not going to be a good day.''}
The Redditor stated that lack of sleep is a trigger, which implied that they anticipate an epileptic episode the following day and are interested in knowing if other members of this community have had this experience. 

Other Redditors expressed concerns about nocturnal seizures and any activity that happens in an unconscious state. Even though unconsciousness is not a top word, it is a prominent concern because it encapsulates nocturnal seizures and any activity that happens when asleep. When discussing nocturnal seizures, we saw Redditors crowd-source information by having shared their experiences to figure out if they had a nocturnal seizure: \emph{``Is this a nocturnal seizure? So around 3 a.m, \underline{I} woke up and felt an electric like feeling and according to my boyfriend,\underline{I} was having convulsions for about two minutes''}
The Redditor expressed uncertainty about their experience and was looking for an answer as to whether or not the symptoms they experienced at night can indeed be nocturnal epilepsy/seizures. After disclosing details of the event, the Redditor concluded their post by reemphasizing the original question: \emph{``How can \underline{I} be sure if this is a nocturnal seizure?''} Even PLWE who knew they experienced nocturnal seizures still asked questions to see if others had similar experiences. \emph{``For anyone with nocturnal seizures, do you remember them? \underline{I've} only had 2 nocturnal seizures. \underline{I} don't remember anything.\underline{I} wake up confused and have wet myself, and for day afterward \underline{I} have a sore body and tongue''}
Here, we observed a Redditor who was aware of their nocturnal seizures and, in turn, asked others who shared similar experiences of not remembering what happened.

\subsubsection{Photosensitivity} 

Even though photosensitivity does not affect the entire population of PLWE, for those PLWE who are affected, the impact can be severe. Many posts about photosensitivity were about how different forms of media could trigger a seizure for PLWE. The results of the LIWC-22 analysis show that the posts on this topic mainly discuss risks and illnesses related to leisure in their lives. With a mean of 2.95, this topic is the highest in leisure. The category of leisure includes keywords such as 'game, fun, play' and other words that indicate leisure activities. In particular, the use of flashing lights and images in various forms of media: \emph{``Incredibles 2 warning if you are light sensitive. We were watching \underline{The Incredibles} 2 with our kids yesterday, which all the \underline{flashing screens} caused my wife to have a seizure''}
This exemplified how flashing scenes in a movie can lead to seizure activity in photosensitive PLWE. It also showed how imperative it is for media content to include flash warnings.

In addition to movies, there is a lot of uncertainty surrounding photosensitive content in video games, video edits, and GIFs. Some Redditors who do not have epilepsy, such as content creators, use r/Epilepsy to better understand the triggers of photosensitive epilepsy so they can avoid triggering a seizure for their viewers: \emph{``I created a video for work with rapid edits. A coworker is worried that it could cause epileptic seizures. Is there any guide/regulations that might show me what is "safe" and "unsafe" in video editing?''}
This presented another use of the r/Epilepsy subreddit: people who do not have epilepsy came to this community to ask questions that they want someone with epilepsy to answer. The Redditor of the previous post chose to ask the affected community (PLWE) how to ensure their video edits were safe for photosensitive PLWE. Like other Redditors who do not live with epilepsy, this subreddit may be viewed as a credible source on matters such as these. We saw this in the following post as well: \emph{``I want to make sure I don't trigger a seizure as I am dating someone with epilepsy. Are there any epilepsy-safe video games?''}
Similarly to the previous post, this Redditor was interested in learning about safer alternatives for the specific media to avoid accidentally triggering a seizure while dating a PLWE.   

\subsection{Treatment and Healthcare experience}  
Healthcare experiences for PLWE vary from person to person, but PLWE came to the r/Epilepsy community to share apprehension about medications, questions about epilepsy, anecdotes about their interaction with their medical professionals, finding new medical providers, and medical testing and diagnostics.
\subsubsection{Medication}  
A daily concern that most PLWE face is the management of anti-epilepsy drugs (AEDs). Redditors posted questions about medications, including side effects, medication dosages, and medication interactions. The posts on this topic mostly used words related to quantity, cause, and acquisition, which aligns with medication dosages and access to medications. This topic has the highest mean in the category of 'acquire' with 1.78. Keywords that belong to this category include 'get' and 'take', see Appendix \ref{section-physical}.
We also observed that the objective of the questions was to learn about the side effects of their prescription and seek validation from others in the community:
\emph{``Experience with Briviact? Does anyone have drowsiness \underline{taking Briviact?}} Or other side effects?''
We learned that this type of general question was not limited to one drug brand, and Redditors asked similar questions about different drug brands (e.g., Briviact, Keppra, etc.). For example, the following Redditor was interested in learning about Keppra's side effects.  

\emph{``I've been on Keppra for about two years and on my current dose of 1750mg twice a day. My seizures have stopped but I've been \underline{getting} a lot of headaches and dizziness. I was wondering if anyone else have had these symptoms after being on a high dose for a while?''}


Similarly, we also saw a group of Redditors that expressed hesitation concerning medication changes, including dosage changes, and asked the community about specific experiences of transitioning from one drug to the other:
\emph{``At what dosage and how long did Xcopri start working? I'm on week five and just started 50mg. I'm not noticing a change yet, but I know I'm still titrating up. Did it \underline{take} the full 12 weeks for you to see an improvement?''}
This Redditor was interested in learning from others at what dosage they started to notice the medication taking effect, which sheds light on another question PLWE might have when starting new medications. We also learned that some PLWE show caution when taking medicine other than AEDs (anti-epileptic drugs), such as antibiotics:  

\emph{``So I'm on Keppra and Klonopin. I was recently prescribed the antibiotic Cipro for an infection, which I have to \\underline{take twice a day.} Should I stagger the times \underline{I take} my AEDs and the antibiotics?
\\ Can \underline{taking all these prescription drugs} at the same time reduce the effectiveness of any of them?''}

This Redditor questioned the effectiveness of antibiotics and AEDs if taken together, illustrating how some PLWE take precautions with taking additional drugs and potential issues that others might face when being prescribed AEDs and non-AEDs.

\subsubsection{Understanding epilepsy}
We discovered that Redditors asked questions to understand different types of epilepsy and their characteristics. Most of these posts were in the form of questions that relate to the LIWC-22 analysis, which shows that Redditors predominantly used words related to cognitive insights (e.g., know, how, think, feel) and lack of cognitive states (e.g., don't have, less). This topic is one of the highest in the category of cognition with a mean of 16.84, see Appendix \ref{section-cognition}.

For example, one of the Redditors asked about different types of seizures: \emph{``I've heard about "partial \underline{absence} seizures" but \underline{I don't know} what it is. Is it a complex partial seizure with loss of consciousness?''}
While it is unclear whether this Redditor is a PLWE or not, they still showed interest in knowing about partial absence seizures. Similarly, a few posts asked others in the community about their epilepsy:
\emph{``Mine are partial onset seizures, perhaps originating in the temporal or frontal lobe, sometimes developing into generalized tonic-clonic seizures. What kind of epilepsy do you have?''}
We learned from this Redditor about one of the possible types of seizures PLWE may experience. 

However, we discovered an interest in asking the community questions about their specific form of epilepsy. Supplementary to asking about epilepsy, we noticed questions about medical diagnosis tests such as magnetic resonance imaging (MRI) and electroencephalograph (EEG) testing: \emph{``Do you have abnormal MRI?. \underline{I know} most of the time  MRI or EEG  comes back normal. Mine showed left hippocampal sclerosis''}
The Redditor disclosed their MRI results; additionally, they inadvertently mentioned a common problem with EEGs and MRIs for some PLWE.

\emph{``Normal EEG? I'm diagnosed with tonic epilepsy. There been times when I have three to four seizures a week. My EEG havent shown abnormal brain activity or epileptic activity. Does it change my diagnosis? \underline{I don't know what to do. I don't }
\underline{know what it means}''}


An EEG is a helpful medical diagnostic tool that measures electrical brain activity, but this Redditor noted that a major caveat is that EEGs can produce normal results if there is no seizure activity when the EEG is conducted. We know from this Redditor that they were diagnosed with tonic epilepsy and have had seizures in the past; however, their EEG showed normal brain activity, which in turn worried them. As a result, a medical provider's role becomes imperative in addressing this problematic issue, as discussed next.  

\subsubsection{Healthcare experience}
Healthcare experience encapsulated all things related to a traditional medical setting, such as appointment scheduling or referrals, interactions with medical providers, and the logistics of medical testing. Redditors discussed their interactions with physicians, which align with the LIWC-22 results that show high use of 3rd person pronouns with a mean of 0.99 and 0.74 in the categories of 'she/he' and 'they', respectively. and high in the Social category with a mean of 163.75, specifically with terms that exemplify Communication such as 'said, say, tell, thank'. See Appendix \ref{section-linguistic}

We noticed several PLWE discussed their experiences and difficulties trying to see the right doctor: \emph{``I had a doctors appointment yesterday for my seizures and she finally \underline{said} she would get me a referral for a neurologist...Is it normal for it to take that long to get a referral?''}
This Redditor noted that they needed a referral from their primary care provider to see a neurologist, which is an issue for people experiencing epilepsy-related symptoms who do not have an established neurologist. However, finding a neurologist can be challenging due to excessive wait times. 
\emph{`` I was referred to a neurologist over 20 weeks ago and I still don't know when the appointment will actually be''}. Having a neurologist is not the same as having the right neurologist; as members of the r/Epilepsy community shared, PLWE who have an established neurologist did not always feel that their concerns were heard: \emph{``After seeing my neurologist for 6 years I'm giving up on \underline{him.} I'm tired of \underline{him} questioning me and not taking my concerns seriously. I have found another doctor who specializes in epilepsy and am going to make an appointment.''}
Unfortunately, some PLWE may experience this Redditor's similar experience.  
The relationship between PLWE and their provider is crucial for proper treatment. As we saw in the r/Epilepsy posts, a lot of community members post questions about others' healthcare experiences as well as share their experiences. 

\subsubsection{Medical Research and Treatments}
Some posts focused on different treatments and research, from brain surgeries to medical studies, which parallels the LIWC-22 analysis that revealed that the posts regarding this topic primarily focused on physical health. This topic is the highest in the category of physical (15.02), and subcategory of health (11.89). Some keywords in these categories include 'medic', 'patients', 'health', 'sick'. See Appendix \ref{section-physical}. The Redditors who were interested in brain surgery were curious about its success: \emph{``Has brain surgery cured anyone's seizures? Just curious if anyone has had brain surgery performed to get rid of their seizures. Was it successful or not?''}
Brain surgery is an option for some PLWE, and people seem to turn to the r/Epilepsy community to ask about others' experiences with these operations. Complimenting questions about brain surgeries, posts also shared research findings in very simple terms: 
\emph{``Study Shows Positive Outcomes For Treating Epilepsy With CBD''}  
Although some of these posts might be from accredited medical research related to epilepsy, many posts did not provide references to the published research.
Even though these posts differ, they are alike in presenting epilepsy concerning medical conditions. Next, we present what we learned about epilepsy on a much more daily basis. 

\subsection{Daily Functions}  
PLWE also discussed how epilepsy affects their everyday activities and experiences that are not unique to PLWE. We noticed the frequency of seizure activity can contribute to a person living with epilepsy's perceived level of independence. We also witnessed posts about PLWE reaching milestones of being seizure-free. In addition, we learned that PLWE posted about finances and lifestyle choices, such as dietary options.  
\subsubsection{Perceived level of Independence}  
We acknowledge that for many Americans, driving is associated with a perception of independence. Unfortunately, many Redditors expressed that they are not allowed to drive because of their seizure activity. Driving laws vary by state, but the consensus is that there is a mandatory period after seizure activity when driving is not allowed. Therefore, for PLWE who live in car-dependent areas, driving can become more of a hindrance to their perception of independence than a freedom. We learned that even after being cleared to drive, some PLWE were not entirely comfortable doing so as some PLWE experience seizures without warning. As one Redditor described, while they are physically able to drive, they are unsure if they are safe to drive again. As a result, multiple posts in r/Epilepsy crowd-sourced advice to overcome hesitation with driving. According to the LIWC-22 results, the posts in this category have the highest means in the Social subcategories of family (0.77) and female (0.95)

\emph{``Anxious about driving? I'm 5 months seizure-free next week. It's legal in my state to drive once you're medically cleared and have been seizure-free for 3 months. I've been afraid of driving ever since my seizure last year. I rarely leave the house unless I either have to or have someone else to pick me up. It's caused relationship issues. I'm afraid I'll never be able to enjoy driving again. Does anyone else experience this? How do you convince yourself that you're safe to drive?''}

We also learned that being seizure-free and regaining driving privileges was a milestone that Redditors wanted to share: \emph{``This has been the first time in two and a half years where I have ever reached the goal of being seizure free for 6 months and I am beyond happy. You will reach your seizure-free goals as well! If I can do it, you can do it too''} Another Redditor emphasized this point with immense excitement.


PLWE expressed that reaching these milestones is not an easy feat. In the case of this Redditor, it took more than two years to reach the milestone of being seizure-free for six months. They expressed happiness over this goal and provided words of encouragement for others who may be in a similar situation.

Unfortunately for some Redditors, they used r/Epilepsy to share their feelings about their seizure-free streak having come to an end: \emph{``After seizure-free months, I've had them come back yesterday and today. I have to wear a sling for a while after my last seizure. Just needed a place to vent. How is everyone else's condition doing?''}
This Redditor made the most of this subreddit by sharing their personal experience and inviting others to share their personal experiences. We learned that these concerns are closely related to PLWE view on finances.
\subsubsection{Finances}  
This topic included factors that can contribute to financial stability, such as jobs, medical insurance, and prescription costs. In terms of jobs, Redditors discussed concerns with finding employment that aligns with their needs. Additionally, posts that focused on medical insurance talked about the pros and cons of higher-paying jobs vs lower-paying jobs with better health benefits. This correlates with our LIWC-22 results that show this topic has the highest mean of 7.75 in the Lifestyle category due to the mainly used words relating to work (mean = 5.08), money (mean = 2.40), see Appendix \ref{section-lifestyle}.

Many PLWE expressed concerns about finding a job as they rely on medical insurance, whether public or private. Private insurance can be obtained through employers, thus making employment with health care benefits a critical factor for people with a chronic illness such as epilepsy. However, for some PLWE, finding a job can be proved to be a daunting task, especially if driving is restricted. As a result, we observed that some PLWE sought advice for finding jobs online: \emph{``Online jobs? I am trying to get back in the swing of everything, but since I can't drive and don't have someone to take me to a job during the days, I have been looking for an online job.''}
As we have previously seen, the concern of driving is prominent, thus forcing some PLWE, such as this Redditor, to look for online jobs. In this case, we saw how the inability to drive posed a restriction.
For PLWE who can drive, the task of switching jobs can also be worrisome in terms of having to switch insurance. The question of prescription and medical provider coverage can arise, as following.

\emph{``My insurance won't cover Vimpat if I switch jobs. I haven't had the interview yet, but I was contacted by a contractor to work for a company and was sent the information about the healthcare insurance. This job would give me \underline{\$10k extra a year} but won't cover the cost of Vimpat at all, any suggestions?''}

Not only is it important to have health benefits from employers, PLWE also discussed trouble with whether or not the insurance provided by employers is sufficient. In the case of the Redditor, they had to decide between a higher salary or the coverage of their medication. The cost of Vimpat and other brand-name drugs can be very costly. Therefore, we noticed that many Redditors posted questions about how to manage prescription costs:  

\emph{``What do you about prescription \underline{cost}? I am switching back from (generic) Keppra to (name brand) Dilantin. \underline{It cost three} \underline{times as much.} What do you do to keep \underline{cost} down?''}

This Redditor questioned how to manage the increase in prescription costs after changing from a generic (non-name brand) to a name-brand medication. While there are generic alternatives to most name brands, there are some instances in which PLWE are prescribed brand-name drugs.  This proved to become a financial concern, especially since, for some, the only option is a name-brand prescription. 

\subsubsection{Lifestyle choices}
We found that some PLWE posted questions about lifestyle choices such as food and alcohol. Also, according to the results of the LIWC-22 analysis, this topic has the second highest mean of 13.06 in the category of physical, with the highest mean in the wellness subcategory with a mean of 1.94. Wellness-relevant words such as healthy, gym, diet were mostly used in this topic. See Appendix \ref{section-physical}.
Most PLWE asked about dietary choices, such as adding or removing foods, and their effect on seizure activities: \emph{``What would you recommend to \underline{reduce seizure symptoms diet wise?} I have a lot of quick foods like TV dinners and some fast food. A lot of water and some fruit drinks. I used to drink coffee every day, and now I've cut back.''}
In addition to food, we noticed the author mentioned a decrease in their coffee consumption as a dietary change. We learned that other Redditors also asked about caffeine as a potential seizure trigger. In addition to discussing caffeine consumption, we learned that many Redditors were interested in learning about alcohol and epilepsy: \emph{``If we stay hydrated while drinking (water with our beer), can we get drunk ever again, or is it a lost cause?''}
In terms of lifestyle choices, we noted several posts inquiring about alcohol. Many AEDs are known to come with an alcohol warning, thus prompting Redditors to ask if there is any alternative. 

\subsection{Seizure Activity}   
There are four phases of epileptic seizures: prodromal, pre-ictal (auras), ictal, and post-ictal; each phase is differentiated by the physical symptoms usually experienced. Posts that discussed seizure activity provided more information about auras, ictal activities, and seizure monitoring and alert.  
\subsubsection{Auras}
The pre-ictal phase, commonly known as an "aura", is generally a warning sign for PLWE of an impending seizure. Even though auras are a type of seizure, most PLWE consider it a predecessor to a major epileptic episode. For some PLWE, auras can be a change in cognitive processes, which aligns with the LIWC-22 analysis that showed this topic to be the highest in the category of Cognition with a mean of 17.25 due to use of  words related to cognitive process, memory, anxiety, and tentativeness found in these posts. See Appendix \ref{section-cognition}. We found that most posts concerning the pre-ictal phases sought validation from others who might experience the same sensation.   

\emph{``Just wondering if anyone can relate to this experience. I'll be sitting watching TV when suddenly have an image of something in my mind. The image comes with a sinking, pit of your stomach-type feeling. A feeling of \underline{dread} with nausea. This lasts about 20 sec to 1 minute max. When the image pops in my head I know the sense of \underline{dread} soon follows. This has happened around five times, just wondering if anyone has auras that are similar?''}
This Redditor shared a detailed account of how their auras are. This post noted that the primary concern of this post was auras as opposed to seizures, previous posts we've seen discussed topics in terms of seizures, while this topic focused mostly on auras. We saw that for some PLWE, there were other symptoms besides seizures, such as auras. 
\subsubsection{Ictal}
The ictal phase is the main seizure event. Since this topic focuses on seizure activities, LIWC-22 analysis also indicated that posts related to this topic exhibited negative tones and negative emotions with a mean of 3.03 and 1.64 respectively. See Appendix \ref{section-affect}. Some PLWE described their personal seizure episodes, while others described side effects that occur during seizure activity,y such as self-injury.

\emph{``Have you ever injured yourself due to a seizure? This Sunday morning I had a seizure, a tonic-clonic. I've injured myself before - I've sliced my face open on a glass panel in a door I fell through, to mention the worst one until now. I fell on my head, cut the flesh around my eye \underline{badly}, broke my glasses (they're practically a part of me) and broke a rib. Anyway how have you injured yourself?  How do you deal with them?''}

Unfortunately, this Redditor experienced multiple forms of self-injury as a result of a tonic-clonic seizure. While we saw various posts that mentioned self-injury, most of these posts specified tongue biting and bruising. Nonetheless, it is important to take into consideration the additional concern of self-injury in addition to seizure activities. In turn, any assistive device or medical alert bracelet can benefit PLWE.  

\subsubsection{Seizure Monitoring and Alert} 
With the advancement of technology, the question of seizure alert devices has become relevant on the subreddit. This corroborates the results of the LIWC-22 which showed that posts on this topic predominantly use words and expressions related to technology (mean of 2.38) and culture (mean of 2.44), exhibiting more positive emotions (mean of 0.94) and curiosity (mean of 1.60). Some keywords that exhibit culture include car and phone. See Appendix \ref{section-culture}. Redditors expressed interest in any technology that can serve to alert loved ones to seizure activity.

\emph{``Emergency alert devices for epileptics? My husband is always close by. Is there any sort of pager or \underline{lifealert type} \underline{device} I could use to let him know I'm about to go down? Some sort of button I could carry with me, and he could have a \underline{receiver}. It'd be cool if he had a \underline{pager} that would tell him my location and to get me. Anything like this out there?''}


This Redditor showed interest in an alert device that can be used when they feel as if they are going down. However, the device that is in question depends on the user being able to know preemptively when a seizure might happen. Unfortunately, for some PLWE, a seizure can happen without any warning. In this case, a medical alert ID can help paramedics and others know how to help the person in need:  

\emph{``My dad has epilepsy and just recently had a breakthrough seizure. I'm looking to get him a bracelet/ID for him to wear that is big enough to have three contact numbers as well as his medical info. What do you guys wear/use? Any recommendations? Thanks in advance!''}

We discovered that not only PLWE were interested in this topic but also were their loved ones, as seen by this post. Since the Redditor is a non-PLWE, they were seeking advice from members of this community. We learned that medical alert IDs can include very helpful information for aiding a PLWE in need, such as emergency contact and medical history, and help with the support of the wearer.
 
\subsection{Support for PLWE} 
While this theme was not the most prominent, it is equally important as it included the topics of support, awareness, and assistance for PLWE with resources that can help support them, such as support groups and service animals. 
\subsubsection{Support for PLWE}
In terms of support for PLWE we observed posts that discussed ways of finding support for epilepsy. For example, the following Redditor was curious to know about epilepsy support groups and the potential therapeutic benefits of joining a support group. Active support groups can allow this Redditor to find others with similar struggles. Support platforms not only aid PLWE, but also help raise awareness and inform others of epilepsy: \emph{``Is anyone involved with support groups involving epilepsy? It's therapeutic finding a group of people with similar frustrations, and I'm hoping to find more active groups with similar struggles.Thanks''}
Posts also shared resources that others living with epilepsy can utilize and gather together and share their own stories. The Redditors expressed how sharing stories can help raise awareness. Additionally, they described how the resource can help loved ones of PLWE understand more about epilepsy. 
\subsubsection{PLWE awareness and assistance}
Moreover, we saw that authors were interested in service animals; some shared stories of witnessing service dogs alert their owners, others inquired about the training of seizure-alert dogs: \emph{``People with Seizure Dogs, did you get them pre-trained? Did you get them and then have them trained? Is it special training or can you train a dog to detect a seizure by being around you enough? I just got a dog, and I'm curious.''}
Seizure dogs are trained in the detection and warning of an impending seizure, therefore prompting some Redditors to wonder if they are able to train their own dogs or if more specialized training is needed. Seizure dogs are one of the ways that PLWE can have support. Posts that mentioned epilepsy awareness discussed charities, awareness walks, t-shirts, and individuals raising awareness: \emph{``Miss Pre-Teen West Fargo Makes Lemonade for Epilepsy Awareness''}. This post exemplified one of the many ways that individuals can help raise epilepsy awareness. In terms of raising awareness the LIWC-22 showed that the posts mostly used words that are associated with moral (mean = 0.25) and social referents (mean = 6.65). See Appendix \ref{section-social}.

In summary, the 15 topics provided insight into identifying five primary areas of discussion that capture the breadth and depth of the concerns shared in the community, from managing unpredictable symptoms and navigating complex treatment pathways to coping with daily life challenges and seeking online social support.
 \section{Discussion} 

In this section, we contextualize our results within the landscape of previous research, discuss technology design opportunities to address the concerns raised in the r/Epilepsy community, and conclude with some limitations and potential future research. 

\subsection{Inter-connectivity of Epilepsy Topics}
Epilepsy is a complicated chronic condition that is difficult to manage. Our findings shed light on the discussion topics and online behaviors in the r/Epilepsy community, which differ from epilepsy-related content on other platforms. While previous research indicates that epilepsy-related posts on other sites, including Facebook and Twitter, are for raising awareness and providing support \cite{meng_social_2017}, we found that PLWE on Reddit choose to self-disclose their conditions, ask questions, seek validation and support, and crowd-source relevant information. This can be due to the fundamental differences in design, affordances, and user populations between Reddit and other social media platforms. Since epilepsy is a stigmatized condition, PLWE might have difficulty disclosing their experience with epilepsy \cite{ni2001relationship}. Since Reddit allows anonymity, we studied a community where PLWE may feel comfortable discussing more freely. 

Our results corroborated previous findings on relevant topics in epilepsy diagnosis and identified less obvious areas of concern. For example, previous work also identified mental health (Topic 1), sleep (Topic 0), and treatment (Topic 4, 12) as relevant concerns for PLWE \cite{fazekas_insights_2021}. Our findings build on this knowledge with areas of concern for the holistic experience of living with epilepsy, including perceived level of independence (topic 14), finances (topic 8), and lifestyle choices (topic 13). Together, our results highlight the prominence of not only epilepsy-related challenges but also challenges concerning daily functions. Hence, we advocate for future researchers and developers interested in designing technologies for PLWE to develop a holistic understanding of their challenges and concerns. 
 

We also found that many topics in our dataset are interconnected, meaning the themes coexist and can indirectly or directly impact each other. For example, r/Epilepsy users noted that seizure activity can impact daily functions, such as driving, which can affect other aspects of life, such as employment \emph{``Any jobs that are remote? I can't drive because of my epilepsy.''} Users also point to a connection between mental health and seizure activity, such as the impact that epilepsy has on mental health \emph{``has anyone else finally had enough of being so angry, depressed, and stressed out?''} and the impact that mental health can have on seizure activity \emph{``my triggers are lack of sleep and stress.''} This interconnectivity further demonstrates that managing and living with epilepsy is incredibly complicated, which needs to be taken into account when designing technologies for PLWE.

In the subsequent sections, we describe opportunities for holistic epilepsy self-management technology design based on an established chronic condition self-management framework \cite{miller_chronic_2015}. This framework is designed to account for the complex and interconnected nature of chronic conditions such as epilepsy, as demonstrated in our results. This framework categorizes self-management systems into three system levels that must be incorporated to maximize wellness -- the intrapersonal system (section \ref{intrapersonal}), the interpersonal-selected support system (section \ref{interpersonal}), and the environmental system (section \ref{environmental}). We describe what each system is and use the topics from our results to propose design opportunities for each system level. Although the systems are presented in separate sections, these features can be implemented into one cohesive epilepsy self-management technology.    


\subsection{Mental and Physical Health and Information Seeking (Intrapersonal System)}
\label{intrapersonal}
The intrapersonal system focuses on individual self-awareness and self-management. It involves assessing emotions and physical conditions, utilizing resources, being aware of potential changes, understanding health literacy, and planning for life interruptions \cite{miller_chronic_2015}. As previously discussed, the themes in our dataset are interactive and interwoven; one example of this includes mental and physical health, which are overlapping topics in our dataset.
This makes sense given that, compared to the general population, a higher proportion of PLWE experience psychiatric comorbidities such as depression and anxiety \cite{gilliam2003psychiatric}. Furthermore, these comorbidities can contribute to an increase in seizure frequency~\cite{jackson2005depression, fazekas_insights_2021}.
We advocate for a multi-faceted approach to epilepsy care for PLWE \cite{kerr2012impact}. This requires managing both mental and physical health in one holistic intervention, which has been shown to decrease depression and epilepsy issues~\cite{sajatovic2016targeted}.
Most epilepsy mobile apps focus on treatment management, including medication adherence, healthcare communication, and seizure tracking \cite{alzamanan_self-management_2021, escoffery_review_2018}. While these functions are essential, integrating features that are tailored for mental health, specifically as a result of epilepsy, can aid successful self-management. We anticipate features inspired by existing mental health applications that users perceive as helpful, including goal-setting, behavior monitoring, and advice and information \cite{thornton2018specific}. Setting goals has been shown to improve the quality of life of chronic condition patients compared to those who do not engage in goal-setting \cite{tabaei2023goal}.
Some tools offer epilepsy logging and tracking~\cite{escoffery_review_2018}, but applications can also include mood-logging features for anxiety attacks or logging of potential triggers to mental distress. These logging features encourage users to consider other ways their health may be impacted and provide insight into the relationship between mental and physical health for PLWE. This creates an opportunity for PLWE to adjust their daily function based on how they feel emotionally and physically, which is a requirement for managing a chronic condition \cite{miller_chronic_2015}. 


Health literacy is another component of the intrapersonal system and is characterized as the knowledge of signs, symptoms, and factors that may exacerbate a chronic condition \cite{miller_chronic_2015}. Research has shown that health literacy impacts health information-seeking behavior \cite{lee2021role, ellis2012role}, which can aid PLWE in learning health information-seeking behavior that can improve self-management. We discovered that across the various topics of discussion, PLWE asked questions and crowd-sourced information. Therefore, future technological design should consider implementing multiple ways that allow PLWE to learn and digest epilepsy-related information. Building off of the previously discussed addition of features to epilepsy self-management applications, developers could design the implementation of gamification and educational features such as educational games and personalized information feeds that cater to the specific needs of each PLWE. Previous research analyzed gamification design and its relation to chronic condition self-management \cite{huang2023use} and how it can be leveraged for chronic condition management \cite{miller2016game}. Our results show the willingness of PLWE to ask questions and seek information, which can be supplemented by educational games that encourage and reward PLWE. As seen in our findings, PLWE sought to understand more about epilepsy in terms of different types of seizures and side effects. Gamification features can include educational games with reward systems (e.g., points and badges) that teach about epilepsy and the different types of seizures. Another design feature to include is a personalized information feed that caters to each user. This information feed can consist of relevant information that we found was discussed most among PLWE, such as lifestyle choices, and treatment and healthcare experience. To further enhance the information feed, it can be tailored to each unique user based on volunteered personal information such as epilepsy type, seizure type, and medications taken. This will allow for a more personalized self-management experience.



\subsection{Online Communities and Peer Support (Interpersonal System)} 
\label{interpersonal}
The interpersonal-selected support system involves communication with family and utilizing selected support persons to aid in self-management. Patient-selected support persons are individuals living with chronic conditions who are selected as designated support persons who can provide emotional and social support \cite{miller_chronic_2015}. In some cases, there is a mutual exchange of emotional and social support if the patient-selected support person also lives with the same or a similar chronic condition. Previous research shows that PLWE want to share information with, and seek information from, other PLWE \cite{Min2023}. Our results showed that PLWE encouraged this mutual exchange in the r/Epilepsy community through inviting others to share their experiences, asking for support and/or information, and seeking validation from other group members. We discovered how PLWE ask questions related to medical and social life.
Even though epilepsy is a stigmatized condition, we saw several posts share very personal (non-identifiable) stories and detailed accounts of life with epilepsy. This highlights the relationship between anonymity and disclosure among PLWE, in which anonymity may facilitate self-disclosure among individuals who experience illness-related embarrassment \cite{rains2014implications}. As such, technology designers can leverage privacy-preserving design to create a safe environment that encourages disclosure among PLWE.

From our results, we also learned that PLWE appreciate other PLWE and the /Epilepsy community, which emphasizes the positive impact that support from online health communities and social media can have on an individual's health \cite{Hong2013, Moorhead2013, Ma2017b}. While subreddits, such as r/Epilepsy, are a great way of connecting Redditors, they lack the intimacy that peer support groups and smaller-sized communities can provide. Peer support groups have gained popularity and have demonstrated their efficacy in chronic condition care \cite{fisher2023peer}. Although the r/Epilepsy community is well-appreciated, it might be challenging for PLWE to establish intimate connections with other PLWE. The r/Epilepsy community continues to grow, which is beneficial for community building but not necessarily for intimate connections. Therefore, designs incorporating smaller peer support groups that emphasize peer-to-peer interactions in existing online communities may further aid PLWE concerning the interpersonal dimension of self-management. As seen from our results, numerous topics are discussed among PLWE. Peer support groups can be formed around various areas of concern we discovered and help PLWE connect with other PLWEs beyond sharing an epilepsy diagnosis. For example, groups may revolve around the different treatments/medications used for epilepsy, such as vagus nerve stimulator (VNS) vs. Keppra vs. surgery, or specific challenges as a result of epilepsy, such as epilepsy and employment. Emphasizing features such as private peer support groups and communities on social media platforms such as Reddit will aid Redditors with epilepsy to find a more intimate, tailored-fit community where they can exchange emotional and social support. Additionally, improving the current flair (tags that relate to their specific topics) \cite{vanclassifying} descriptions and usage in the r/Epilepsy subreddit will allow PLWE to find and filter posts with relatable information, such as medication, gender, age range, and seizure type, which will help navigate a rather large online community.

\subsection{Healthcare Community and Public Awareness (Environmental System)} 
\label{environmental}
The environmental systems extend beyond the individual to the broader healthcare community. It emphasizes effective communication and collaboration with healthcare providers and resources to optimize daily performance, well-being, and health management\cite{miller_chronic_2015}. In our results, we discovered that PLWE posted about their interactions with healthcare professionals and their eagerness to seek and receive support from people beyond the confines of online communities by raising epilepsy awareness. However, when discussing their interactions with healthcare providers, we found that some PLWE are less than satisfied with these interactions. Our results also highlight how posts discussing healthcare have a high use of words that exemplify communication. This confirms previous research of effective communication as essential to improving health outcomes \cite{simon2021complex}. Alongside these posts and the observed medical information-seeking online behavior, our results signify how online communities may be a first resort for PLWE to seek medical information, especially about medications. This could result from the average wait times for a neurologist or epileptologist, which we discovered was challenging for some PLWE. While we saw one PLWE waiting over 20 weeks, the average wait time for new appointments is 35 business days which is still longer than other specialties \cite{craft2013}. In addition to challenges accessing medical providers, our results highlight the value of lived experience. Several posts asked other PLWE about their experience concerning medical concerns such as medication, diagnosis, and treatment. While online health communities are great resources for social support and information seeking, there exists an opportunity for a better approach to ensuring that PLWE can access both lived experiences and medically accurate information. Also, some topics discussed, especially the ones that include medical information, will benefit from information validation and verification checks by medical professionals to avoid misinformation \cite{trethewey2020strategies}. Medical professional volunteers could potentially moderate these groups or answer some of the common questions where peer knowledge may not be enough \cite{huh2016lessons}. For some known medical misinformation, machine learning models can help detect and add warning labels to the posts. However, possible misclassification from AI models could lead to mistrust and needs to be implemented carefully \cite{jian2000foundations}. 

Prior literature has argued for the necessity of improving public awareness of epilepsy, suggesting that historical and cultural depictions could have influenced this lack of awareness while also enhancing misconceptions and stigma \cite{Kobau2021, de2010epilepsy, MCCAGH20091,krauss2000scarlet}. As we witnessed PLWE discussing different methods for raising epilepsy awareness, we recommend that future technologies be designed to foster heightened public awareness at various levels, considering cultural factors. This could be implemented within the community of PLWE, extending to broader organizational levels like hospitals and school systems and ultimately reaching the general public. Previous research on technologies for individuals with various chronic conditions, such as migraines and HIV, has similarly highlighted the significance of promoting heightened awareness and reducing misconceptions and stigma \cite{Jacoby2002, Park2015_migraine, maestre2020hiv}. It has been studied that prior education of people, including individuals within PLWE's closed social network and the public, would help provide appropriate support to PLWE \cite{Price2004, Gadow1982}. Numerous approaches have been recognized for enhancing awareness and diminishing misconceptions and stigma. These include using campaigns, advertisements, lectures, public service announcements, and educational materials \cite{Chakraborty2021}. From our results, some PLWE showed interest in service dogs. Previous research has shown that service dog handlers with invisible disabilities report experiencing more discrimination than other service dog handlers \cite{mills2017invisible}. Therefore in addition to epilepsy awareness, education and awareness about resources, aids and assistance such as service dogs can help PLWE. 

\subsection{Limitations and Future Research}  
\label{limitations}
Despite the unique insights from studying Reddit, we acknowledge that this approach has limitations to consider. 
For example, even though Reddit requires users to be at least 13 years old~\cite{proferes_studying_2021} and some reports suggest it is most popular among males who are 18 to 34~\cite{proferes_studying_2021}, we cannot verify users' actual ages. Furthermore, although  r/Epilepsy is a popular subreddit with more than 50,000 users - more than other epilepsy-focused communities at the time of data collection, such as PatientsLikeMe, which has 3,073 users \cite{de_la_loge_patientslikeme_2016} - our conclusions remain limited to the users we studied and should not be generalized to others. Even though we had a large sample of users in our dataset, the anonymous nature of Reddit means we cannot look for specific sub-populations of PLWE with unique concerns. 
Future research may analyze other epilepsy-based forums or deploy different methods, such as interviews, to discover if and how our findings vary for specific sub-populations or extend beyond r/Epilepsy. 
Additionally, researchers can apply our approaches to understand areas of concern for other populations of interest, such as older adults and PLWE, who are caretakers for individuals who do not have epilepsy.

\section{Conclusion} 
Epilepsy is one of the most common, life-altering neurological conditions, and PLWE rely on social media as a resource. We sought to provide a deeper description of the topics and challenges PLWE seek support for online. To do this, we analyzed 23,944 posts collected from the r/Epilepsy subreddit and analyzed them with topic modeling and qualitative inquiry. Our results point to 15 primary areas of concern, which we group into five themes: Symptoms and Triggers, Treatment and Healthcare Experience, Daily Functions, Seizure Activity, and Support for PLWE. We discuss the implications of our findings for mental and physical health, information seeking, online communities and peer support, healthcare, and public awareness and highlight design implications that leverage the themes' inter-connective nature for technical interventions to aid in epilepsy self-management. 

\section{Author Contribution}
Jessica Y. Medina and Afsaneh Razi contributed to the conception and design of the study. Jessica Y. Medina, Jordyn Young, and Aehong Min were responsible for formal analysis and drafting the manuscript. Patrick C. Shih and Wendy R. Miller provided feedback for intellectual content. Afsaneh Razi oversaw the project from conception to final draft. All authors contributed to the review and editing of the manuscript, approved the final version to be published, and agree to be accountable for all aspects of the work.

\bibliographystyle{ACM-Reference-Format}
\bibliography{references.bib,zotero_references.bib}

\appendix

    

\clearpage

\clearpage
\section{LIWC-22 Results}

\subsection{Affect}
\label{section-affect} 
\begin{landscape}
\begin{table}[H]
\resizebox{\columnwidth}{!}{%
\begin{tabular}{|l|l|l|l|l|l|l|l|l|l|l|l|l|}
\hline
 &  &  & Affect & tone\_pos & tone\_neg & emotion & emo\_pos & emo\_neg & emo\_anx & emo\_anger & emo\_sad & swear \\ \hline
Topic & Code & N & Mean & Mean & Mean & Mean & Mean & Mean & Mean & Mean & Mean & Mean \\ \hline
Sleep/Nocturnal & 0 & 2865 & 4.23 & 1.78 & 2.27 & 1.86 & 0.40 & 1.39 & 0.40 & 0.11 & 0.10 & 0.14 \\ \hline
Mental health and memory & 1 & 4370 & 5.98 & 2.76 & 2.86 & \cellcolor[HTML]{739FEB}2.76 & 0.86 & \cellcolor[HTML]{739FEB}1.71 & 0.43 & \cellcolor[HTML]{739FEB}0.31 & \cellcolor[HTML]{739FEB}0.33 & \cellcolor[HTML]{739FEB}0.24 \\ \hline
Photosensitivity & 2 & 324 & \cellcolor[HTML]{739FEB}6.05 & 3.27 & 2.71 & 2.09 & 0.79 & 1.28 & 0.27 & 0.05 & 0.05 & 0.06 \\ \hline
Symptoms (Ictal) & 3 & 639 & 5.21 & 1.95 & \cellcolor[HTML]{739FEB}3.03 & 2.19 & 0.49 & 1.64 & 0.20 & 0.16 & 0.07 & 0.21 \\ \hline
Medication & 4 & 2620 & 4.26 & 1.95 & 2.08 & 1.81 & 0.37 & 1.26 & 0.40 & 0.13 & 0.16 & 0.06 \\ \hline
Seizure monitoring/alert & 5 & 201 & 5.81 & 3.82 & 1.97 & 2.52 & \cellcolor[HTML]{739FEB}0.94 & 1.57 & 0.14 & 0.02 & 0.07 & 0.02 \\ \hline
Support for PLWE & 6 & 108 & 3.60 & 2.58 & \cellcolor[HTML]{FD6864}0.77 & \cellcolor[HTML]{FD6864}0.70 & \cellcolor[HTML]{FD6864}0.16 & \cellcolor[HTML]{FD6864}0.30 & \cellcolor[HTML]{FD6864}0.02 & \cellcolor[HTML]{FD6864}0.01 & \cellcolor[HTML]{FD6864}0.00 & \cellcolor[HTML]{FD6864}0.00 \\ \hline
Healthcare & 7 & 1437 & 3.81 & 2.30 & 1.36 & 1.33 & 0.50 & 0.73 & 0.27 & 0.09 & 0.07 & 0.11 \\ \hline
Finances & 8 & 539 & 3.95 & 2.40 & 1.46 & 1.14 & 0.52 & 0.57 & 0.18 & 0.07 & 0.03 & 0.24 \\ \hline
Assisting PLWE & 9 & 1271 & 5.81 & \cellcolor[HTML]{739FEB}4.32 & 1.42 & 1.68 & 0.79 & 0.86 & 0.08 & 0.07 & 0.10 & 0.06 \\ \hline
Symptoms (Auras) & 10 & 2183 & 3.90 & \cellcolor[HTML]{FD6864}1.36 & 2.39 & 1.89 & 0.27 & 1.52 & \cellcolor[HTML]{739FEB}0.83 & 0.06 & 0.07 & 0.07 \\ \hline
Epilepsy the illness & 11 & 1886 & \cellcolor[HTML]{FD6864}3.58 & 1.73 & 1.74 & 1.43 & 0.34 & 1.00 & 0.31 & 0.08 & 0.09 & 0.03 \\ \hline
Medical research/treatment & 12 & 764 & 4.39 & 3.18 & 1.16 & 1.02 & 0.37 & 0.60 & 0.18 & \cellcolor[HTML]{FD6864}0.01 & 0.03 & 0.01 \\ \hline
Lifestyle choices & 13 & 318 & 4.10 & 2.52 & 1.52 & 1.22 & 0.44 & 0.75 & 0.12 & 0.04 & 0.06 & 0.04 \\ \hline
Freedom and restrictions & 14 & 4420 & 4.97 & 2.85 & 1.99 & 1.96 & 0.70 & 1.18 & 0.55 & 0.09 & 0.12 & 0.08 \\ \hline
\rowcolor[HTML]{FFCE93} 
 & Total & 23945 & 4.72 & 2.41 & 2.13 & 1.92 & 0.56 & 1.26 & 0.41 & 0.13 & 0.14 & 0.12 \\ \hline
\end{tabular}%
}

\label{tab:my-table}
\end{table}
\end{landscape}

\newpage
\subsection{Cognition}  
\label{section-cognition}
\begin{landscape}
\begin{table}[H]
\resizebox{\columnwidth}{!}{%
\begin{tabular}{|l|l|l|l|l|l|l|l|l|l|l|l|l|}
\hline
 &  &  & Cognition & allnone & cogproc & insight & cause & discrep & tentat & certitude & differ & memory \\ \hline
Topic & Code & N & Mean & Mean & Mean & Mean & Mean & Mean & Mean & Mean & Mean & Mean \\ \hline
Sleep/Nocturnal & 0 & 2865 & 13.39 & 1.25 & 12.07 & 2.81 & 1.44 & 1.51 & 3.42 & 0.56 & 3.65 & 0.30 \\ \hline
Mental health and memory & 1 & 4370 & 16.86 & \cellcolor[HTML]{739FEB}1.35 & 15.43 & 4.06 & 2.09 & 2.03 & 3.81 & \cellcolor[HTML]{739FEB}0.75 & 3.93 & 0.49 \\ \hline
Photosensitivity & 2 & 324 & 11.06 & 0.68 & 10.38 & 3.06 & 1.99 & 1.30 & 3.15 & 0.19 & 2.50 & 0.03 \\ \hline
Symptoms (Ictal) & 3 & 639 & 12.37 & 1.03 & 11.28 & \cellcolor[HTML]{FD6864}2.25 & 1.64 & 1.32 & 3.54 & 0.46 & 3.57 & 0.10 \\ \hline
Medication & 4 & 2620 & 15.23 & 0.99 & 14.18 & 3.38 & 2.70 & 1.53 & 4.36 & 0.45 & 4.16 & 0.13 \\ \hline
Seizure monitoring/alert & 5 & 201 & 10.99 & 0.65 & 10.30 & 2.59 & 1.85 & 1.23 & 2.95 & 0.24 & 2.41 & 0.23 \\ \hline
Support for PLWE & 6 & 108 & \cellcolor[HTML]{FD6864}9.04 & 0.53 & \cellcolor[HTML]{FD6864}8.51 & 4.04 & \cellcolor[HTML]{FD6864}1.10 & \cellcolor[HTML]{FD6864}0.68 & \cellcolor[HTML]{FD6864}1.83 & \cellcolor[HTML]{FD6864}0.17 & \cellcolor[HTML]{FD6864}1.64 & \cellcolor[HTML]{FD6864}0.00 \\ \hline
Healthcare & 7 & 1437 & 14.35 & 0.98 & 13.31 & 3.35 & 1.81 & 1.99 & 3.76 & 0.45 & 3.63 & 0.11 \\ \hline
Finances & 8 & 539 & 13.07 & 0.81 & 12.17 & 2.52 & 1.73 & \cellcolor[HTML]{739FEB}2.12 & 3.32 & 0.31 & 3.62 & 0.08 \\ \hline
Assisting PLWE & 9 & 1271 & 12.88 & 0.66 & 12.14 & 3.55 & 1.66 & 1.78 & 3.16 & 0.37 & 2.83 & 0.12 \\ \hline
Symptoms (Auras) & 10 & 2183 & \cellcolor[HTML]{739FEB}17.25 & 1.25 & \cellcolor[HTML]{739FEB}15.90 & 4.26 & 1.41 & 1.41 & \cellcolor[HTML]{739FEB}4.92 & \cellcolor[HTML]{FFFFFF}0.74 & \cellcolor[HTML]{739FEB}4.72 & \cellcolor[HTML]{739FEB}0.53 \\ \hline
Epilepsy the illness & 11 & 1886 & 16.84 & 1.02 & 15.73 & \cellcolor[HTML]{739FEB}4.30 & 2.20 & 1.61 & 4.83 & 0.45 & 4.11 & 0.17 \\ \hline
Medical research/treatment & 12 & 764 & 12.17 & \cellcolor[HTML]{FD6864}0.46 & 11.66 & 2.83 & 2.48 & 1.40 & 3.34 & 0.26 & 2.56 & 0.16 \\ \hline
Lifestyle choices & 13 & 318 & 16.02 & 1.02 & 14.97 & 3.30 & \cellcolor[HTML]{739FEB}2.65 & 2.06 & 4.56 & 0.50 & 4.30 & 0.05 \\ \hline
Freedom and restrictions & 14 & 4420 & 13.47 & 1.12 & 12.26 & 2.83 & 1.93 & 1.72 & 3.22 & 0.53 & 3.42 & 0.11 \\ \hline
\rowcolor[HTML]{FFCE93} 
 & Total & 23945 & 14.78 & 1.09 & 13.62 & 3.42 & 1.95 & 1.69 & 3.81 & 0.54 & 3.74 & 0.25 \\ \hline
\end{tabular}%
}

\label{tab:my-table}
\end{table}
\end{landscape}

\newpage 
\subsection{Culture} 
\label{section-culture}
\begin{landscape}
\begin{table}[H]
\resizebox{\columnwidth}{!}{%
\begin{tabular}{|l|l|l|l|l|l|l|}
\hline
 &  &  & Culture & politic & ethnicity & tech \\ \hline
Topic & Code & N & Mean & Mean & Mean & Mean \\ \hline
Sleep/Nocturnal & 0 & 2865 & 0.23 & 0.02 & 0.01 & 0.21 \\ \hline
Mental health and memory & 1 & 4370 & 0.32 & 0.04 & 0.02 & 0.26 \\ \hline
Photosensitivity & 2 & 324 & 1.72 & 0.09 & 0.05 & 1.58 \\ \hline
Symptoms (Ictal) & 3 & 639 & 0.42 & 0.05 & \cellcolor[HTML]{FD6864}0.00 & 0.36 \\ \hline
Medication & 4 & 2620 & \cellcolor[HTML]{FD6864}0.17 & 0.05 & 0.01 & \cellcolor[HTML]{FD6864}0.11 \\ \hline
Seizure monitoring/alert & 5 & 201 & \cellcolor[HTML]{739FEB}2.44 & 0.03 & 0.04 & \cellcolor[HTML]{739FEB}2.38 \\ \hline
Support for PLWE & 6 & 108 & 1.20 & 0.12 & \cellcolor[HTML]{739FEB}0.22 & 0.86 \\ \hline
Healthcare & 7 & 1437 & 0.52 & 0.09 & 0.03 & 0.40 \\ \hline
Finances & 8 & 539 & 0.99 & \cellcolor[HTML]{739FEB}0.44 & 0.05 & 0.50 \\ \hline
Assisting PLWE & 9 & 1271 & 1.52 & 0.18 & 0.11 & 1.23 \\ \hline
Symptoms (Auras) & 10 & 2183 & 0.18 & \cellcolor[HTML]{FD6864}0.01 & 0.01 & 0.16 \\ \hline
Epilepsy the illness & 11 & 1886 & 0.31 & 0.03 & 0.02 & 0.26 \\ \hline
Medical research/treatment & 12 & 764 & 1.11 & 0.19 & 0.08 & 0.83 \\ \hline
Lifestyle choices & 13 & 318 & 0.20 & 0.03 & 0.01 & 0.15 \\ \hline
Freedom and restrictions & 14 & 4420 & 0.37 & 0.06 & 0.01 & 0.30 \\ \hline
\rowcolor[HTML]{FFCE93} 
 & Total & 23945 & 0.45 & 0.06 & 0.02 & 0.36 \\ \hline
\end{tabular}%
}

\label{tab:my-table}
\end{table}
\end{landscape}

\newpage 
\subsection{Drives} 
\label{section-drives}
\begin{landscape}
\begin{table}[H]
\resizebox{\columnwidth}{!}{%
\begin{tabular}{|l|l|l|l|l|l|l|}
\hline
 &  &  & Drives & affiliation & achieve & power \\ \hline
Topic & Code & N & Mean & Mean & Mean & Mean \\ \hline
Sleep/Nocturnal & 0 & 2865 & 2.07 & 0.98 & 0.67 & 0.43 \\ \hline
Mental health and memory & 1 & 4370 & 2.83 & 1.32 & 0.95 & 0.57 \\ \hline
Photosensitivity & 2 & 324 & 2.82 & 1.31 & \cellcolor[HTML]{FD6864}0.45 & 1.10 \\ \hline
Symptoms (Ictal) & 3 & 639 & 2.00 & 0.72 & 0.66 & 0.68 \\ \hline
Medication & 4 & 2620 & 1.85 & 0.59 & 0.94 & \cellcolor[HTML]{FD6864}0.33 \\ \hline
Seizure monitoring/alert & 5 & 201 & 2.48 & 1.15 & 0.95 & 0.38 \\ \hline
Support for PLWE & 6 & 108 & 3.74 & 1.36 & \cellcolor[HTML]{FFFFFF}0.46 & \cellcolor[HTML]{739FEB}2.01 \\ \hline
Healthcare & 7 & 1437 & 2.32 & 0.93 & 0.83 & 0.60 \\ \hline
Finances & 8 & 539 & 4.54 & 1.32 & \cellcolor[HTML]{739FEB}1.92 & 1.32 \\ \hline
Assisting PLWE & 9 & 1271 & \cellcolor[HTML]{739FEB}4.98 & \cellcolor[HTML]{739FEB}2.80 & 1.28 & 1.03 \\ \hline
Symptoms (Auras) & 10 & 2183 & \cellcolor[HTML]{FD6864}1.39 & \cellcolor[HTML]{FD6864}0.44 & 0.54 & 0.43 \\ \hline
Epilepsy the illness & 11 & 1886 & 1.61 & 0.67 & 0.52 & 0.48 \\ \hline
Medical research/treatment & 12 & 764 & 3.04 & 0.67 & 1.48 & 0.97 \\ \hline
Lifestyle choices & 13 & 318 & 3.16 & 0.87 & 1.88 & 0.68 \\ \hline
Freedom and restrictions & 14 & 4420 & 3.28 & 1.49 & 1.05 & 0.79 \\ \hline
\rowcolor[HTML]{FFCE93} 
 & Total & 23945 & 2.60 & 1.11 & 0.91 & 0.62 \\ \hline
\end{tabular}%
}

\label{tab:my-table}
\end{table}
\end{landscape}

\newpage 
\subsection{Lifestyle}
\label{section-lifestyle}
\begin{landscape}
\begin{table}[H]
\resizebox{\columnwidth}{!}{%
\begin{tabular}{|l|l|l|l|l|l|l|l|l|}
\hline
 &  &  & Lifestyle & leisure & home & work & money & relig \\ \hline
Topic & Code & N & Mean & Mean & Mean & Mean & Mean & Mean \\ \hline
Sleep/Nocturnal & 0 & 2865 & 1.91 & 0.30 & \cellcolor[HTML]{739FEB}0.50 & 0.92 & 0.15 & 0.05 \\ \hline
Mental health and memory & 1 & 4370 & 1.86 & 0.38 & 0.13 & 1.14 & 0.16 & 0.07 \\ \hline
Photosensitivity & 2 & 324 & 4.82 & \cellcolor[HTML]{739FEB}2.96 & 0.14 & 1.39 & 0.21 & 0.16 \\ \hline
Symptoms (Ictal) & 3 & 639 & 1.47 & 0.43 & 0.26 & 0.66 & 0.07 & 0.06 \\ \hline
Medication & 4 & 2620 & 1.24 & \cellcolor[HTML]{FD6864}0.08 & \cellcolor[HTML]{FD6864}0.04 & 0.93 & 0.17 & 0.02 \\ \hline
Seizure monitoring/alert & 5 & 201 & 3.08 & 0.50 & 0.15 & 1.54 & 0.80 & 0.11 \\ \hline
Support for PLWE & 6 & 108 & 5.68 & 0.49 & 0.35 & 3.99 & 0.68 & \cellcolor[HTML]{739FEB}0.18 \\ \hline
Healthcare & 7 & 1437 & 2.98 & 0.17 & 0.09 & 2.33 & 0.37 & 0.05 \\ \hline
Finances & 8 & 539 & \cellcolor[HTML]{739FEB}7.75 & 0.22 & 0.18 & \cellcolor[HTML]{739FEB}5.08 & \cellcolor[HTML]{739FEB}2.40 & 0.05 \\ \hline
Assisting PLWE & 9 & 1271 & 4.08 & 0.39 & 0.09 & 2.99 & 0.60 & 0.08 \\ \hline
Symptoms (Auras) & 10 & 2183 & \cellcolor[HTML]{FD6864}1.05 & 0.23 & 0.17 & \cellcolor[HTML]{FD6864}0.57 & \cellcolor[HTML]{FD6864}0.05 & 0.04 \\ \hline
Epilepsy the illness & 11 & 1886 & 1.36 & 0.12 & 0.07 & 1.07 & 0.08 & 0.03 \\ \hline
Medical research/treatment & 12 & 764 & 2.91 & 0.20 & 0.18 & 2.37 & 0.14 & 0.04 \\ \hline
Lifestyle choices & 13 & 318 & 1.42 & 0.28 & 0.12 & 0.89 & 0.12 & \cellcolor[HTML]{FD6864}0.02 \\ \hline
Freedom and restrictions & 14 & 4420 & 2.19 & 0.32 & 0.19 & 1.43 & 0.23 & 0.05 \\ \hline
\rowcolor[HTML]{FFCE93} 
 & Total & 23945 & 2.15 & 0.31 & 0.18 & 1.39 & 0.25 & 0.05 \\ \hline
\end{tabular}%
}

\label{tab:my-table}
\end{table}
\end{landscape}

\newpage 
\subsection{Linguistic}
\label{section-linguistic}
\begin{landscape}
\begin{table}[H]
\resizebox{\columnwidth}{!}{%
\begin{tabular}{|l|l|l|l|l|l|l|l|l|l|l|l|l|l|l|l|l|l|l|l|l|l|l|l|}
\hline
 &  &  & Linguistic & function & prounoun & ppron & i & we & you & shehe & they & ipron & det & article & number & prep & auxverb & adverb & conj & negate & verb & adj & quantity \\ \hline
Topic & Code & N & Mean & Mean & Mean & Mean & Mean & Mean & Mean & Mean & Mean & Mean & Mean & Mean & Mean & Mean & Mean & Mean & Mean & Mean & Mean & Mean & Mean \\ \hline
Sleep/Nocturnal & 0 & 2865 & \cellcolor[HTML]{FFFFFF}73.49 & 58.42 & 16.51 & \cellcolor[HTML]{739FEB}11.59 & \cellcolor[HTML]{739FEB}8.92 & 0.23 & 0.66 & \cellcolor[HTML]{739FEB}1.10 & 0.58 & 4.92 & \cellcolor[HTML]{739FEB}13.49 & 5.59 & 2.56 & 13.00 & 9.80 & 6.58 & 8.00 & 1.57 & 18.54 & 5.81 & 5.20 \\ \hline
Mental health and memory & 1 & 4370 & \cellcolor[HTML]{FFFFFF}71.65 & 56.75 & 17.13 & 11.08 & 7.90 & 0.33 & 1.34 & 0.72 & 0.59 & 6.05 & 11.99 & 4.49 & 1.75 & 12.23 & 9.80 & 7.04 & 7.08 & \cellcolor[HTML]{739FEB}1.63 & \cellcolor[HTML]{739FEB}18.97 & 5.84 & 4.07 \\ \hline
Photosensitivity & 2 & 324 & \cellcolor[HTML]{FFFFFF}51.38 & 38.79 & 8.97 & 5.39 & 3.02 & 0.36 & 1.01 & 0.44 & 0.39 & 3.58 & 9.31 & 4.36 & 2.90 & 10.96 & 6.63 & 3.50 & 4.03 & 0.93 & 12.51 & 4.89 & 2.47 \\ \hline
Symptoms (Ictal) & 3 & 639 & \cellcolor[HTML]{FFFFFF}67.37 & 53.32 & 14.90 & 9.68 & 7.22 & 0.22 & 0.94 & 0.67 & 0.50 & 5.22 & 13.08 & 5.40 & 1.45 & 12.60 & 8.57 & 5.85 & 6.87 & 1.15 & 16.12 & 6.44 & 3.97 \\ \hline
Medication & 4 & 2620 & \cellcolor[HTML]{FFFFFF}69.37 & 54.61 & 14.62 & 9.06 & 6.77 & 0.15 & 1.05 & 0.50 & 0.48 & 5.56 & 11.90 & 4.52 & 2.91 & 12.46 & 10.11 & 6.30 & 7.48 & 1.14 & 18.11 & 5.60 & \cellcolor[HTML]{739FEB}5.67 \\ \hline
Seizure monitoring/alert & 5 & 201 & \cellcolor[HTML]{FFFFFF}51.15 & 39.18 & 10.12 & 6.30 & 3.25 & 0.29 & \cellcolor[HTML]{739FEB}1.90 & 0.40 & 0.33 & 3.82 & 10.09 & 4.10 & \cellcolor[HTML]{739FEB}4.10 & 10.32 & 6.27 & 3.85 & 4.14 & 0.62 & 13.00 & 3.98 & 2.67 \\ \hline
Support for PLWE & 6 & 108 & \cellcolor[HTML]{FD6864}39.70 & \cellcolor[HTML]{FD6864}30.41 & \cellcolor[HTML]{FD6864}7.17 & \cellcolor[HTML]{FD6864}4.77 & \cellcolor[HTML]{FD6864}2.23 & 0.23 & \cellcolor[HTML]{FD6864}0.71 & \cellcolor[HTML]{739FEB}1.10 & \cellcolor[HTML]{FD6864}0.29 & \cellcolor[HTML]{FD6864}2.40 & \cellcolor[HTML]{FD6864}7.20 & \cellcolor[HTML]{FD6864}3.69 & \cellcolor[HTML]{FD6864}1.42 & \cellcolor[HTML]{FD6864}10.31 & \cellcolor[HTML]{FD6864}4.48 & \cellcolor[HTML]{FD6864}1.88 & \cellcolor[HTML]{FD6864}3.43 & \cellcolor[HTML]{FD6864}0.42 & \cellcolor[HTML]{FD6864}9.39 & \cellcolor[HTML]{FD6864}3.48 & \cellcolor[HTML]{FD6864}1.46 \\ \hline
Healthcare & 7 & 1437 & \cellcolor[HTML]{FFFFFF}69.01 & 54.78 & 15.31 & 10.41 & 7.17 & 0.30 & 1.05 & 0.99 & \cellcolor[HTML]{739FEB}0.74 & 4.89 & 13.33 & \cellcolor[HTML]{739FEB}5.82 & 2.13 & 12.44 & 9.78 & 5.72 & 6.41 & 1.47 & 18.62 & 5.16 & 4.19 \\ \hline
Finances & 8 & 539 & \cellcolor[HTML]{FFFFFF}64.27 & 51.79 & 13.85 & 9.24 & 6.41 & 0.36 & 1.21 & 0.48 & 0.66 & 4.61 & 11.71 & 4.90 & 1.97 & 13.12 & 9.28 & 4.85 & 6.01 & 1.47 & 16.40 & 5.21 & 3.55 \\ \hline
Assisting PLWE & 9 & 1271 & \cellcolor[HTML]{FFFFFF}58.64 & 46.14 & 11.41 & 7.02 & 3.82 & \cellcolor[HTML]{739FEB}0.54 & 1.45 & 0.46 & 0.51 & 4.39 & 11.11 & 4.87 & 1.92 & \cellcolor[HTML]{739FEB}13.22 & 7.60 & 4.00 & 5.14 & 0.79 & 14.59 & 5.02 & 3.01 \\ \hline
Symptoms (Auras) & 10 & 2183 & \cellcolor[HTML]{739FEB}73.70 & \cellcolor[HTML]{739FEB}58.74 & \cellcolor[HTML]{739FEB}16.82 & 10.51 & 8.56 & \cellcolor[HTML]{FD6864}0.10 & 0.75 & 0.40 & 0.62 & \cellcolor[HTML]{739FEB}6.31 & 12.97 & 5.22 & 1.76 & 12.39 & 10.03 & \cellcolor[HTML]{739FEB}7.30 & \cellcolor[HTML]{739FEB}8.50 & 1.51 & 18.54 & \cellcolor[HTML]{739FEB}6.47 & 4.41 \\ \hline
Epilepsy the illness & 11 & 1886 & \cellcolor[HTML]{FFFFFF}66.07 & 53.81 & 14.26 & 8.83 & 6.05 & 0.19 & 1.05 & 0.74 & 0.62 & 5.43 & 12.06 & 4.70 & 2.10 & 11.77 & \cellcolor[HTML]{739FEB}10.58 & 6.02 & 7.05 & 1.21 & 17.06 & 5.91 & 4.03 \\ \hline
Medical research/treatment & 12 & 764 & \cellcolor[HTML]{FFFFFF}51.63 & 40.63 & 8.14 & 4.78 & 2.66 & 0.19 & 1.09 & 0.28 & 0.38 & 3.36 & 9.25 & 4.39 & 2.33 & 12.63 & 7.39 & 3.62 & 4.55 & 0.73 & 12.48 & 5.36 & 2.96 \\ \hline
Lifestyle choices & 13 & 318 & \cellcolor[HTML]{FFFFFF}66.14 & 52.05 & 13.80 & 8.52 & 6.19 & 0.17 & 1.47 & \cellcolor[HTML]{FD6864}0.26 & 0.34 & 5.27 & 11.21 & 4.63 & 1.92 & 11.02 & 9.87 & 5.65 & 7.17 & 1.22 & 17.61 & 5.99 & 4.86 \\ \hline
Freedom and restrictions & 14 & 4420 & \cellcolor[HTML]{FFFFFF}68.95 & 54.70 & 14.78 & 10.24 & 7.22 & 0.32 & 0.95 & 1.06 & 0.53 & 4.54 & 12.58 & 4.91 & 2.91 & 12.76 & 10.07 & 6.21 & 6.86 & 1.36 & 17.92 & 5.70 & 5.09 \\ \hline
\rowcolor[HTML]{FFCE93} 
 & Total & 23945 & 68.44 & 54.31 & 15.00 & 9.82 & 7.06 & 0.27 & 1.04 & 0.75 & 0.56 & 5.18 & 12.25 & 4.91 & 2.30 & 12.47 & 9.63 & 6.15 & 6.99 & 1.35 & 17.65 & 5.71 & 4.46 \\ \hline
\end{tabular}%
}

\label{tab:my-table}
\end{table}
\end{landscape}

\newpage 
\subsection{Physical} 
\label{section-physical}
\begin{landscape}
\begin{table}[H]
\resizebox{\columnwidth}{!}{%
\begin{tabular}{|l|l|l|l|l|l|l|l|l|l|l|l|l|l|l|l|l|l|l|l|l|l|}
\hline
 &  &  & Physical & health & illness & wellness & mental & substances & sexual & food & death & need & want & acquire & lack & fulfill & fatigue & reward & risk & curiosity & allure \\ \hline
Topic & Code & N & Mean & Mean & Mean & Mean & Mean & Mean & Mean & Mean & Mean & Mean & Mean & Mean & Mean & Mean & Mean & Mean & Mean & Mean & Mean \\ \hline
Sleep/Nocturnal & 0 & 2865 & 7.23 & \cellcolor[HTML]{FD6864}4.21 & 2.83 & 0.06 & 0.12 & 0.13 & 0.04 & 0.37 & 0.07 & 0.42 & 0.22 & 1.15 & 0.26 & 0.13 & \cellcolor[HTML]{739FEB}0.36 & \cellcolor[HTML]{FD6864}0.02 & 0.31 & \cellcolor[HTML]{FD6864}0.29 & 7.40 \\ \hline
Mental health and memory & 1 & 4370 & 5.84 & 4.53 & 2.76 & 0.18 & 0.42 & 0.21 & 0.07 & 0.25 & 0.14 & 0.49 & \cellcolor[HTML]{739FEB}0.51 & 0.96 & 0.15 & 0.08 & 0.15 & 0.05 & 0.33 & 0.37 & \cellcolor[HTML]{739FEB}8.14 \\ \hline
Photosensitivity & 2 & 324 & 10.25 & 9.27 & \cellcolor[HTML]{739FEB}8.14 & 0.28 & 0.17 & 0.07 & 0.01 & 0.10 & 0.34 & 0.28 & 0.21 & \cellcolor[HTML]{FD6864}0.46 & 0.11 & 0.05 & 0.04 & 0.11 & \cellcolor[HTML]{739FEB}1.45 & 0.35 & 5.52 \\ \hline
Symptoms (Ictal) & 3 & 639 & 9.80 & 4.83 & 3.12 & 0.21 & 0.09 & 0.06 & 0.01 & 0.39 & 0.14 & 0.41 & 0.20 & 1.01 & 0.11 & 0.11 & 0.13 & \cellcolor[HTML]{FD6864}0.02 & 0.84 & 0.34 & 6.38 \\ \hline
Medication & 4 & 2620 & 7.09 & 5.71 & \cellcolor[HTML]{FD6864}1.77 & 0.26 & 0.41 & 0.29 & 0.20 & 0.28 & 0.05 & 0.31 & 0.29 & \cellcolor[HTML]{739FEB}1.78 & 0.15 & 0.09 & 0.26 & 0.07 & 0.40 & 0.48 & 6.69 \\ \hline
Seizure monitoring/alert & 5 & 201 & 7.80 & 6.64 & 5.05 & 0.18 & 0.19 & 0.22 & \cellcolor[HTML]{FD6864}0.00 & \cellcolor[HTML]{FD6864}0.08 & 0.16 & 0.30 & 0.22 & 0.63 & 0.08 & 0.03 & 0.01 & 0.08 & 1.04 & \cellcolor[HTML]{739FEB}1.60 & 4.59 \\ \hline
Support for PLWE & 6 & 108 & 11.40 & 8.26 & 7.50 & 0.26 & \cellcolor[HTML]{FD6864}0.00 & \cellcolor[HTML]{FD6864}0.00 & \cellcolor[HTML]{FD6864}0.00 & 0.87 & \cellcolor[HTML]{739FEB}0.84 & 0.23 & \cellcolor[HTML]{FD6864}0.03 & 0.64 & \cellcolor[HTML]{FD6864}0.06 & \cellcolor[HTML]{FD6864}0.02 & \cellcolor[HTML]{FD6864}0.00 & 0.08 & 0.90 & 0.38 & \cellcolor[HTML]{FD6864}2.54 \\ \hline
Healthcare & 7 & 1437 & 6.71 & 5.50 & 2.54 & 0.09 & 0.14 & 0.25 & 0.06 & 0.21 & 0.05 & 0.61 & 0.39 & 1.18 & 0.17 & 0.10 & 0.10 & 0.08 & 0.31 & 0.39 & 6.72 \\ \hline
Finances & 8 & 539 & \cellcolor[HTML]{FD6864}5.60 & 4.79 & 2.44 & 0.06 & 0.02 & 0.22 & 0.02 & 0.12 & 0.05 & \cellcolor[HTML]{739FEB}0.81 & 0.30 & 0.94 & 0.18 & \cellcolor[HTML]{739FEB}0.15 & 0.04 & 0.19 & 1.15 & 0.48 & 7.40 \\ \hline
Assisting PLWE & 9 & 1271 & 8.15 & 7.11 & 5.30 & 0.40 & 0.24 & 0.31 & 0.03 & 0.13 & 0.05 & 0.29 & 0.37 & 0.62 & 0.14 & 0.10 & 0.03 & 0.22 & \cellcolor[HTML]{FD6864}0.24 & 0.86 & 5.66 \\ \hline
Symptoms (Auras) & 10 & 2183 & 6.41 & \cellcolor[HTML]{FD6864}4.21 & 2.51 & \cellcolor[HTML]{FD6864}0.05 & 0.38 & 0.07 & 0.03 & 0.23 & 0.06 & 0.21 & 0.18 & 0.95 & 0.18 & 0.12 & 0.16 & \cellcolor[HTML]{FD6864}0.02 & 0.26 & 0.70 & 7.38 \\ \hline
Epilepsy the illness & 11 & 1886 & 9.97 & 8.49 & 5.76 & 0.12 & \cellcolor[HTML]{739FEB}0.40 & 0.21 & 0.06 & 0.22 & \cellcolor[HTML]{FD6864}0.03 & \cellcolor[HTML]{FD6864}0.18 & 0.20 & 0.71 & \cellcolor[HTML]{739FEB}0.39 & 0.09 & 0.14 & \cellcolor[HTML]{FD6864}0.02 & 0.28 & 0.69 & 5.97 \\ \hline
Medical research/treatment & 12 & 764 & \cellcolor[HTML]{739FEB}15.02 & \cellcolor[HTML]{739FEB}11.89 & \cellcolor[HTML]{FFFFFF}7.96 & 0.41 & 0.18 & 0.68 & \cellcolor[HTML]{739FEB}0.55 & 0.17 & 0.22 & 0.26 & 0.30 & 0.59 & 0.11 & 0.06 & 0.01 & \cellcolor[HTML]{739FEB}0.26 & 0.88 & 0.86 & 4.09 \\ \hline
Lifestyle choices & 13 & 318 & 13.06 & 6.84 & 2.95 & \cellcolor[HTML]{739FEB}1.94 & 0.08 & \cellcolor[HTML]{739FEB}1.80 & 0.01 & \cellcolor[HTML]{739FEB}4.74 & 0.08 & 0.51 & 0.29 & 1.08 & 0.32 & 0.13 & 0.22 & 0.22 & 0.54 & 0.60 & 6.59 \\ \hline
Freedom and restrictions & 14 & 4420 & 7.19 & 5.84 & 4.08 & 0.18 & 0.15 & 0.25 & 0.06 & 0.20 & 0.11 & 0.50 & 0.37 & 1.18 & 0.21 & 0.12 & 0.10 & 0.07 & 0.47 & 0.31 & 7.72 \\ \hline
\rowcolor[HTML]{FFCE93} 
 & Total & 23945 & 7.53 & 5.72 & 3.52 & 0.20 & 0.26 & 0.24 & 0.08 & 0.30 & 0.10 & 0.40 & 0.33 & 1.07 & 0.19 & 0.10 & 0.16 & 0.07 & 0.42 & 0.48 & 7.06 \\ \hline
\end{tabular}%
}

\label{tab:my-table}
\end{table}
\end{landscape}

\newpage 
\subsection{Social} 
\label{section-social}
\begin{landscape}
\begin{table}[H]
\resizebox{\columnwidth}{!}{%
\begin{tabular}{|l|l|l|l|l|l|l|l|l|l|l|l|l|l|l|}
\hline
 &  &  & Social & socbehav & prosocial & polite & conflict & moral & comm & socrefs & family & friend & female & male \\ \hline
Topic & Code & N & Mean & Mean & Mean & Mean & Mean & Mean & Mean & Mean & Mean & Mean & Mean & Mean \\ \hline
Sleep/Nocturnal & 0 & 2865 & 6.58 & 2.44 & 0.53 & 0.28 & 0.10 & 0.07 & 1.46 & 4.04 & 0.38 & 0.13 & 0.68 & 0.90 \\ \hline
Mental health and memory & 1 & 4370 & 8.90 & 3.53 & 0.88 & 0.33 & 0.25 & 0.18 & 1.78 & 5.05 & 0.24 & \cellcolor[HTML]{739FEB}0.18 & 0.50 & 0.67 \\ \hline
Photosensitivity & 2 & 324 & 9.05 & 4.21 & 1.11 & 0.37 & 0.31 & 0.12 & 2.01 & 4.71 & 0.27 & 0.08 & 0.41 & 1.03 \\ \hline
Symptoms (Ictal) & 3 & 639 & 6.42 & 2.26 & 0.46 & 0.21 & 0.21 & 0.10 & 1.18 & 4.05 & 0.17 & 0.05 & 0.43 & 0.64 \\ \hline
Medication & 4 & 2620 & 6.13 & 2.14 & 0.68 & 0.40 & \cellcolor[HTML]{FD6864}0.06 & \cellcolor[HTML]{FD6864}0.05 & 1.27 & 3.92 & 0.18 & \cellcolor[HTML]{FD6864}0.02 & 0.35 & 0.43 \\ \hline
Seizure monitoring/alert & 5 & 201 & 7.65 & 2.68 & 0.61 & 0.32 & 0.07 & 0.13 & 1.26 & 4.79 & 0.30 & 0.08 & 0.34 & 0.55 \\ \hline
Support for PLWE & 6 & 108 & 10.75 & 4.03 & 1.77 & 0.21 & 0.16 & \cellcolor[HTML]{739FEB}0.25 & 1.37 & \cellcolor[HTML]{739FEB}6.65 & 0.40 & 0.12 & 0.69 & \cellcolor[HTML]{739FEB}1.95 \\ \hline
Healthcare & 7 & 1437 & 9.20 & 4.40 & 0.92 & 0.42 & 0.14 & 0.14 & \cellcolor[HTML]{739FEB}2.81 & 4.72 & 0.24 & 0.05 & 0.59 & 0.74 \\ \hline
Finances & 8 & 539 & 9.02 & 3.80 & 1.23 & 0.40 & \cellcolor[HTML]{739FEB}0.37 & 0.13 & 1.73 & 4.82 & 0.23 & 0.05 & 0.38 & 0.54 \\ \hline
Assisting PLWE & 9 & 1271 & \cellcolor[HTML]{739FEB}12.13 & \cellcolor[HTML]{739FEB}5.58 & \cellcolor[HTML]{739FEB}2.06 & \cellcolor[HTML]{739FEB}0.65 & 0.23 & 0.17 & 2.22 & 6.36 & 0.29 & 0.12 & 0.48 & 0.43 \\ \hline
Symptoms (Auras) & 10 & 2183 & \cellcolor[HTML]{FD6864}5.31 & \cellcolor[HTML]{FD6864}2.05 & \cellcolor[HTML]{FD6864}0.36 & 0.22 & 0.11 & 0.06 & 1.36 & \cellcolor[HTML]{FD6864}3.14 & 0.13 & 0.05 & 0.25 & \cellcolor[HTML]{FD6864}0.37 \\ \hline
Epilepsy the illness & 11 & 1886 & 7.11 & 2.46 & 0.57 & 0.45 & 0.09 & 0.08 & 1.52 & 4.53 & 0.24 & 0.05 & 0.49 & 0.57 \\ \hline
Medical research/treatment & 12 & 764 & 7.40 & 2.25 & 0.66 & \cellcolor[HTML]{FD6864}0.20 & 0.09 & 0.06 & \cellcolor[HTML]{FD6864}1.05 & 4.96 & 0.62 & \cellcolor[HTML]{FD6864}0.02 & 0.35 & 0.38 \\ \hline
Lifestyle choices & 13 & 318 & 6.34 & 2.19 & 0.74 & 0.25 & 0.07 & \cellcolor[HTML]{FD6864}0.05 & 1.06 & 3.99 & \cellcolor[HTML]{FD6864}0.08 & 0.09 & \cellcolor[HTML]{FD6864}0.14 & 0.48 \\ \hline
Freedom and restrictions & 14 & 4420 & 8.40 & 2.85 & 0.90 & 0.35 & 0.15 & 0.09 & 1.46 & 5.25 & \cellcolor[HTML]{739FEB}0.77 & 0.11 & \cellcolor[HTML]{739FEB}0.95 & 0.94 \\ \hline
\rowcolor[HTML]{FFCE93} 
 & Total & 23945 & 7.80 & 2.98 & 0.80 & 0.35 & 0.15 & 0.11 & 1.60 & 4.63 & 0.35 & 0.09 & 0.55 & 0.67 \\ \hline
\end{tabular}%
}

\label{tab:my-table}
\end{table}
\end{landscape}

\newpage 
\subsection{Word Count} 
\label{section-wordcount}
\begin{landscape}
\begin{table}[]
\resizebox{\columnwidth}{!}{%
\begin{tabular}{|l|l|l|l|l|l|l|l|l|l|l|}
\hline
 &  &  & WC & Analytic & Clout & Authentic & Tone & WPS & BigWords & Dic \\ \hline
Topic & Code & N & Mean & Mean & Mean & Mean & Mean & Mean & Mean & Mean \\ \hline
Sleep/Nocturnal & 0 & 2865 & \cellcolor[HTML]{739FEB}203.46 & 32.56 & 14.57 & 85.38 & 22.44 & 17.37 & \cellcolor[HTML]{FD6864}15.59 & \cellcolor[HTML]{739FEB}91.90 \\ \hline
Mental health and memory & 1 & 4370 & 159.15 & \cellcolor[HTML]{FD6864}29.41 & 20.87 & 71.41 & 28.60 & 16.29 & 16.95 & 89.80 \\ \hline
Photosensitivity & 2 & 324 & 57.27 & 39.59 & 41.31 & 42.53 & 35.58 & 16.50 & 27.20 & 82.80 \\ \hline
Symptoms (Ictal) & 3 & 639 & 138.98 & 46.35 & 20.61 & 73.61 & 20.47 & 15.11 & 17.26 & 87.41 \\ \hline
Medication & 4 & 2620 & 112.69 & 59.00 & 19.24 & 74.65 & 27.53 & 15.09 & 19.42 & 87.15 \\ \hline
Seizure monitoring/alert & 5 & 201 & 43.31 & 26.92 & 43.08 & \cellcolor[HTML]{FD6864}36.24 & 40.31 & \cellcolor[HTML]{739FEB}17.46 & 25.07 & 78.14 \\ \hline
Support for PLWE & 6 & 108 & \cellcolor[HTML]{FD6864}21.94 & 35.32 & \cellcolor[HTML]{739FEB}53.22 & 39.00 & 38.60 & 15.31 & 30.58 & \cellcolor[HTML]{FD6864}72.73 \\ \hline
Healthcare & 7 & 1437 & 163.75 & 64.36 & 21.26 & 69.09 & 35.13 & 13.62 & 18.97 & 88.72 \\ \hline
Finances & 8 & 539 & 120.60 & 37.09 & 27.55 & 60.20 & 37.77 & 14.92 & 21.68 & 87.66 \\ \hline
Assisting PLWE & 9 & 1271 & 93.99 & 37.06 & 46.09 & 44.05 & \cellcolor[HTML]{739FEB}51.32 & 15.88 & 26.10 & 85.46 \\ \hline
Symptoms (Auras) & 10 & 2183 & 202.85 & 62.58 & \cellcolor[HTML]{FD6864}9.63 & \cellcolor[HTML]{739FEB}89.90 & \cellcolor[HTML]{FD6864}18.16 & 14.67 & 17.73 & 90.66 \\ \hline
Epilepsy the illness & 11 & 1886 & 120.57 & 41.72 & 23.97 & 69.62 & 28.34 & 15.84 & 23.59 & 87.72 \\ \hline
Medical research/treatment & 12 & 764 & 90.91 & 35.75 & 43.21 & 42.30 & 40.74 & 15.27 & \cellcolor[HTML]{739FEB}31.04 & 81.55 \\ \hline
Lifestyle choices & 13 & 318 & 97.12 & 61.59 & 23.85 & 60.92 & 36.13 & 12.48 & 20.19 & 87.90 \\ \hline
Freedom and restrictions & 14 & 4420 & 145.14 & \cellcolor[HTML]{739FEB}71.28 & 25.00 & 72.88 & 35.51 & \cellcolor[HTML]{FD6864}9.66 & 18.72 & 90.28 \\ \hline
\rowcolor[HTML]{FFCE93} 
 & Total & 23945 & 147.17 & 37.15 & 22.78 & 71.57 & 30.36 & 15.98 & 19.46 & 88.80 \\ \hline
\end{tabular}%
}
\label{tab:my-table}
\end{table}
\end{landscape}

\end{document}